\documentclass[useAMS,usenatbib]{mn2e}

\input{psfig.sty}
\usepackage{graphicx}
\usepackage{epsfig}
\usepackage{amssymb}
\usepackage{amsmath}
\usepackage{amsfonts}
\usepackage{txfonts}
\usepackage{multirow}
\usepackage{verbatim}

\title[Non-Gaussian error bars in galaxy surveys -- II]{Non-Gaussian error bars in galaxy surveys -- II}
\author[Joachim Harnois-D\'{e}raps and Ue-Li Pen]{Joachim Harnois-D\'{e}raps$^{1,2}$ 
\thanks{E-mail: jharno@cita.utoronto.ca} and Ue-Li Pen$^{1}$ \thanks{E-mail: pen@cita.utoronto.ca} \\
$^{1}$Canadian Institute for Theoretical Astrophysics, University of
Toronto, M5S 3H8, Canada\\
$^{2}$Department of Physics, University of Toronto, M5S 1A7, Ontario,  Canada}

\begin{document}

\date{\today}

\pagerange{\pageref{firstpage}--\pageref{lastpage}} \pubyear{2011}

\maketitle

\label{firstpage}

\begin{abstract}
Estimating the uncertainty on the matter power spectrum internally (i.e. directly from the data) 
is made challenging by the simple fact that galaxy surveys offer at most a few independent samples. 
In addition, surveys have non-trivial geometries, which make the interpretation of the observations even trickier,
but the uncertainty can nevertheless be worked out within the Gaussian approximation. 
With the recent realization that Gaussian treatments of the power spectrum lead to biased error bars about the dilation of the 
baryonic acoustic oscillation scale, efforts are being directed towards developing non-Gaussian analyses, mainly from N-body simulations so far.
Unfortunately, there is currently no way to tell how the non-Gaussian features observed in the simulations  compare to those of the real Universe,
and it  is generally hard to tell at what level of accuracy the N-body simulations can model 
complicated non-linear effects such as mode coupling and galaxy bias.
We propose in this paper a novel method that aims at measuring non-Gaussian error bars on the matter power spectrum  
directly from galaxy survey data. We utilize known symmetries of the 4-point function, 
Wiener filtering and principal component analysis to estimate the full  covariance matrix from only  four independent fields with minimal prior assumptions.  
We assess the quality of the estimated covariance matrix with a  measurement of the Fisher information content in the amplitude of the power spectrum.
 With the noise filtering techniques and only four fields, we are able to recover the results obtained from a large $N=200$ sample to within $20$ per cent, for $k \le1.0 h\mbox{Mpc}^{-1}$.
We further provide error bars on Fisher information and on the best-fitting parameters, 
and identify which parts of the non-Gaussian features are the hardest to extract. 
Finally, we provide a prescription to extract a noise-filtered, non-Gaussian, covariance matrix from a handful of fields in the presence of a survey selection function.
\end{abstract}

\begin{keywords}
Large scale  structure of Universe --  Dark matter  --  Distance Scale -- Cosmology :  Observations -- Methods: data analysis
\end{keywords}



\section{Introduction}
\label{sec:intro}


The matter power spectrum contains a wealth of information about a number of cosmological parameters,
and measuring its amplitude with per cent level precision has become one of the main task of modern cosmology
 \citep[ Pan-STARRS\footnote{\tt http://pan-starrs.ifa.hawaii.edu/}, DES\footnote{\tt https://www.darkenergysurvey.org/}]{2000AJ....120.1579Y,2003astro.ph..6581C,2007AAS...21113229S,  2010MNRAS.401.1429D, 2012arXiv1211.0310L, 2009ApJ...691..241B}.
Cosmologists are especially interested in the detection of the Baryonic Acoustic Oscillation (BAO) scale,
which allows to measure the evolution of the dark energy equation of state $w(z)$
 \citep{Eisenstein:2005su, 2006A&A...449..891H, 2006PhRvD..74l3507T, 2007MNRAS.381.1053P,  2011MNRAS.415.2892B,
2012arXiv1203.6594A}.

Estimating the mean power spectrum from a galaxy survey is a challenging task, as one needs to incorporate 
the survey mask, model the redshift distortions, estimate the galaxy bias, etc. 
For this purpose, many data analyses follow the prescriptions of \citet{1994ApJ...426...23F} (FKP)   
or the Pseudo Karhunen-Lo\`{e}ve \citep{1996ApJ...465...34V}(PKL hereafter), which provide unbiased estimates of the underlying power spectrum, as long as the observe field is Gaussian in nature.
When these methods are applied on a non-Gaussian field, however, the power spectrum estimator is no longer optimal, and the error about it is biased \citep{2006PhRvD..74l3507T}.

As first discussed in \citet{1999MNRAS.308.1179M} and \citet{2005MNRAS.360L..82R}, Non-Gaussian effects on power spectrum measurements can be quite large; 
for instance, the Fisher information content in the matter power spectrum saturates in the trans-linear regime, which causes the number of degrees of freedom
to deviate from the Gaussian prediction by up to three orders of magnitude. 
The obvious questions to ask, then, are `How do non-Gaussian features propagate on to physical parameters, like the BAO scale, redshift space distortions, neutrino masses, etc.?' and 
`How large is this bias in actual data analyses, as opposed to simulations?'
As a partial answer to the first question, it was shown from a large ensemble of N-body simulations that the Gaussian estimator, acting on non-Gaussian fields, 
produces error bars on the BAO dilation scale \citep{2008PhRvD..77l3540P} that differ from full Gaussian case  by up to 15 per cent  \citep{2012MNRAS.419.2949N}. 

The first paper of this series, \citet{2012MNRAS.423.2288H} (hereafter HDP1) 
addresses the second issue, and measure how large is the bias in the presence of a selection function.
Starting with the 2dFGRS  survey selection function as a study case, and modelling the non-Gaussian features from 200 N-body simulations,
it was found that the difference between Gaussian and non-Gaussian error bars on the power spectrum is {\it enhanced} by the presence of a non-trivial survey geometry\footnote{These error bars are unbiased, but unfortunately not optimal. 
As a matter of fact, the method still suffers from one of the inconvenience of the FKP formalism, 
namely that the convolution with the survey selection function effectively correlates the error bars.
Removing this effect could be done by combining the methodology of HDP1 and that of PKL,
but such a task is beyond the scope of this paper.}.
The 15 per cent bias observed by \citet{2012MNRAS.419.2949N} is, in that sense, a {\it lower bound} on the actual bias that exists in current treatments of the data.

At this point, one could object that the sizes of the biases we are discussing here are very small, 
and that analyses with error bars robust to with 20 per cent are still in excellent shape and
rather robust. However, the story reads differently in the context of dark energy, 
where the final goal of the global international effort is to minimize the error bars about $w(z)$
by performing a succession of experiments with increasing accuracy and resolution.
In the end, a 20 per cent bias on the BAO scale has a quite large impact on the dark energy `figure-of-merit',
and removing this effect is the main goal of this series of paper.

Generally, the onset of non-Gaussianities can be understood from asymmetries that develops in the matter fields
subjected to gravitation, starting at the smallest scales and working their way up to larger scales  \citep{2002PhR...367....1B}.
In Fourier space, Gaussian fields can be completely described by their power spectrum, 
whereas non-Gaussian fields also store information in higher moments.
For instance, the non-linear dynamics that describe the scales with $k > 0.5 h \mbox{Mpc}^{-1}$ tend to couple the Fourier modes of the power spectrum, 
which effectively correlates the measurements. 
This correlation was indeed found from very large samples of N-body simulations \citep{2009ApJ...700..479T},
and act as to lower the number of degrees of freedom in a power spectrum measurement.

One approach that was thought to minimizes these complications consists in excluding most of the non-linear scales, as proposed in \citet{Seo:2003pu}.
However, this cuts out some of the BAO wiggles, thereby reducing our accuracy on the measured BAO scale.
In addition, it is plausible that non-Gaussian features due to the non-linear dynamics, mask, and using simple Gaussian estimators
on non-linear fields interact such as to impact scales as large as $k \sim 0.2 h\mbox{Mpc}$, as hinted by HDP1. 
Optimal analyses must therefore probe the signal that resides in the trans-linear and non-linear scales, 
and construct the power spectrum estimators based 
on known non-Gaussian properties of the measured fields.

Characterization of the non-Gaussian features in the uncertainty about the matter power spectrum  was recently
attempted in HDP1, in which deviations from Gaussian calculations were parameterized by simple fitting functions,
whose best fitting parameters were found from sampling $200$ simulations of dark matter particles.
This was a first step in closing the gap that exists between data analyses and simulations, but is incomplete in many aspects.
First, it does not address the questions of cosmology and redshift dependence, halo bias nor of shot noise, which surely affect  the best-fitting parameters,
and completely overlooks the fact that observations are performed in redshift space. 
All these effects are crucial to understand in order to develop a self-consistent prescription with which we can perform non-Gaussian analyses in data.
Second, and perhaps more importantly, this approach assumes that the non-Gaussian features observed in N-body simulations are unbiased representations of 
those that exist in our Universe. 
Such a correspondence is assumed in any {\it external} error analyses, 
which involve mock catalogues typically constructed either from N-body simulations or from semi-analytical techniques such as
 Log-Normal transforms \citep{1991MNRAS.248....1C} or PThalos  \citep{2002MNRAS.329..629S}. 
In that aspect, the FKP and PKL prescriptions are advantageous since they provide  {\it internal} estimates of the error bars,
hence avoid this issue completely.

%
%

In this second paper (HDP2 hereafter), we set out to determine how well can we possibly measure 
these non-Gaussian features internally from a galaxy survey, including simple cases of survey masks,
with minimal prior assumption. 
In that way, we are avoiding the complications caused by
 the cosmology and  redshift dependence of the non-Gaussian features, and that of using 
incorrect non-Gaussian modelling of the fields. 
  In many aspects, this question  boils down to the problem of estimating error bars using a small number of measurements, with minimal external assumptions.
In such low statistics environment, it has been shown that the shrinkage estimation method \citep{2008MNRAS.389..766P}
can minimize the variance between a measured covariance matrix and a {\it target}, or reference matrix.
Also important to mention is the attempt by \citet{2006MNRAS.371.1188H} to improve the numerical convergence of small samples 
by bootstrap resampling and re-weighting sub-volumes of each realizations, an approach that was unfortunately mined down by the effect of beat coupling  \citep{2006MNRAS.371.1205R}. 

With the parameterization proposed in HDP1, our approach has the advantage to provide physical insights on the non-Gaussian dynamics.
In addition, it turns out that the basic symmetries of the four-point function of the density field found in HDP1 allow us to increase the
number of internal independent subsamples by a large amount, with only
a few quantifiable and generic Bayesian priors.  
In particular, it was shown that the contributions to the power spectrum covariance matrix consist of two parts:  
the Gaussian {\it zero-lag} part, plus a broad, smooth, non-local, non-Gaussian component. 
In this picture, both the Gaussian and non-Gaussian contributions can be accurately estimated even within a single density field, 
because many independent configurations contribute. 
However, any attempt to estimate them {\it simultaneously}, i.e. without differentiating their distinct singular nature, 
results in large sample variance.  

This paper takes advantage of this reduced variance to optimize the measurement of the covariance matrix from only four N-body realizations.
In such  low statistics, the noise becomes dominant over a large dynamical range, hence we propose and compare a number of noise reduction techniques.
In order to gain physical insights, we pay special attention to provide error bars on each of the non-Gaussian parameters and
therefore track down the dominant source of noise in non-Gaussian fields.
To quantify the accuracy of the method, we compare and calibrate our results with a larger sample of 200 N-body realizations,
and use the Fisher information about the power spectrum amplitude as a metric of the performance of each noise reduction techniques.

The issue of accuracy is entangled with a major aspect common to most non-Gaussian analyses: 
they require a measurement of the full power spectrum covariance matrix, which is noisy by nature. 
For instance, an accurate measurement of $N^2$ matrix elements generally requires much more then $N$ realizations;
it is arguable that $N^2$ independent measurements could be enough, but this statement generally depends on the final measurement to be carried
and on the required level of accuracy that is sought.
Early numerical calculations were performed on 400 realizations \citep{2005MNRAS.360L..82R}, 
while  \citet{2009ApJ...700..479T} have performed as many as 5000 full N-body
simulations. \citet{2012MNRAS.419.2949N} have opted for fewer realizations, 
but complemented the measurements with noise reduction techniques, basically a principal component decomposition  \citep[see][for another example]{2009MNRAS.396...19N}.
In any case, it is often unclear how many simulations are required to reach convergence; 
in this paper, we deploy strategies such as bootstrap resampling to assess the degree of precision of our measurements.

We first review in section \ref{sec:bkg} the formalism and the theoretical background relevant for the
estimations of the matter power spectrum and its covariance matrix, with special emphases on the dual nature and symmetries
of the four point function, and on the dominant source of noise in our non-Gaussian parameterization.
We then explain how to improve the measurement of  $C(k,k')$ in section \ref{sec:noiseweight}
with three noise-reduction techniques, and compare the results with a straightforward bootstrap resampling.
In section \ref{sec:fisher}  we measure the Fisher information of  the noise filtered covariance matrices,
and compare our results against the large sample. 
Finally, we describe in section \ref{sec:recipe} a recipe to extract the non-Gaussian features of the power spectrum covariance 
in the presence of a survey selection function, in a low statistics environment, thus completely merging the techniques
developed here with those of HDP1. 
This takes us significantly closer to a stage where we can perform non-Gaussian analyses of actual data.  
Conclusions and discussions are presented in the last section.

\section{Estimation of $C(k,k')$: non-Gaussian parameterization}
\label{sec:bkg}

\subsection{Power spectrum estimator}
\label{subsec:power_spectrum}

In the ideal situation that exists only in N-body simulations --  periodic boundary conditions and
no survey selection function effect -- the matter power spectrum $P({\bf k})$ can be obtained in an unbiased way from
Fourier transforms of the density contrast. The latter is defined as $\delta({\bf x}) \equiv \rho({\bf x})/\bar{\rho} - 1$,
where $\rho({\bf x})$ is the density field and  $\bar{\rho}$ its mean. Namely,
\begin{eqnarray}
        \langle \delta({\bf k}) \delta^{*}({\bf k'})\rangle = (2\pi)^{3}P({\bf k})\delta_{D}({\bf k}-{\bf k'})
\end{eqnarray}
The Dirac delta function enforces the two scales to be identical, and the bracket corresponds to a volume (or ensemble) average.
We refer the reader to HDP1 for details on our simulation suite, and briefly mention here that each of the $N=200$
realization is obtained by evolving $256^3$ particles with {\small CUBEP$^3$M} \citep{2012arXiv1208.5098H} down to a redshift of $z=0.5$, with cosmological parameters 
consistent with the WMAP+BAO+SN five years data release \citep{2009ApJS..180..330K}.
Simulated dark matter particles are assigned onto a grid following  a `cloud in cell' interpolation scheme \citep{1981csup.book.....H},
which is deconvolved in the power spectrum estimation.
The isotropic power spectrum $P(k)$ is finally obtained by taking the average over the solid angle.

\begin{figure}
  \begin{center}
  \epsfig{file=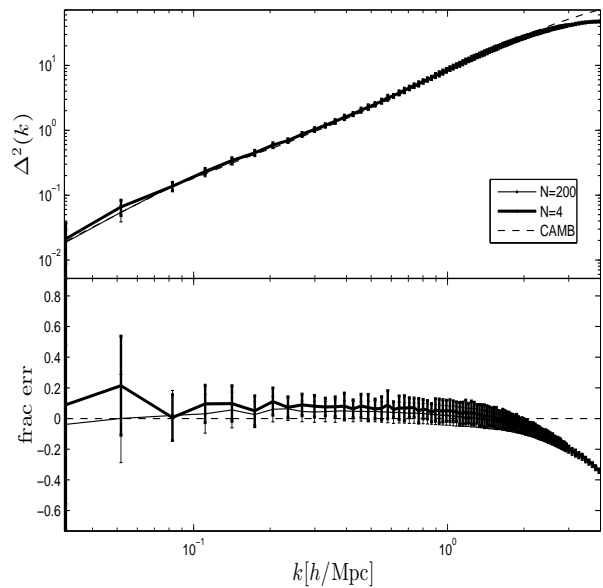,width=0.45\textwidth,height=0.45\textwidth}
  \caption{ ({\it top}:) Dimensionless power spectrum at $z=0.5$, estimated from our template of $200$ N-body simulations (thin line) and from our $4$ `measurements' (thick line). 
  The error bars are the sample standard deviation. Throughout this paper, we represent the $N=4$ and $N=200$ samples with thick and thin solid lines respectively,
  unless otherwise mentioned in the legend or the caption. 
  ({\it bottom}:) Fractional error with respect to the non-linear predictions of {\small HALOFIT} \citep{2003MNRAS.341.1311S}.}
    \label{fig:power_new}
  \end{center}
\end{figure}

Observations depart from this ideal environment: they must incorporate a survey selection function and zero pad the boundary of the survey 
when constructing the power spectrum estimator, plus the galaxy positions are obtained in redshift space. 
We shall return to this framework in section \ref{sec:recipe}, but for now, let focus our attention 
on simulation results.
To quantify the accuracy of our measurements of $P(k)$ and its covariance matrix in the context of low statistics,
we construct a small sample by randomly selecting  $N=4$ realizations among the $200$;
we hereafter refer to these two samples as the $N=200$ and the $N=4$ samples.
Fig. \ref{fig:power_new} shows the power spectrum as measured in these two samples, 
with a comparison to the non-linear predictions of {\small HALOFIT} \citep{2003MNRAS.341.1311S}.
We present the results in the dimensionless form, defined as $\Delta^{2}(k) = k^{3}P(k)/(2 \pi^2)$ in order
to expose the scales that are in the trans-linear regime (loosely defined as the scales where $0.1 < \Delta^{2}(k) < 1.0$)
and those in the non-linear regime (with $ \Delta^{2}(k) > 1.0$).

The simulations show a $\sim 5$ per cent positive bias compared to the predictions, a known systematic 
effect of the N-body code that happens when the simulator is started at very early redshifts.
Starting later would remove this bias, but the covariance matrix, which is the focus of the current paper,
becomes less accurate. This is an unfortunate trade-off that will be avoided in future production runs
with the advent of an initial condition generator based on second order perturbation theory.
The plotted error bars are smaller in the $N=4$ sample, a clear example of bias on the error bars:
the estimate on the variance is poorly determined, as expected from such low statistics,
and the error {\it on} the $N=4$ error bar is quite large.
The mean $P(k)$ measured in the two samples, however, agree at the few per cent level. 
Following HDP1, we extract from this figure the regime of confidence,
inside of which the simulated structures are well resolved, and exclude any information
coming from $k$-modes beyond; in this case, scale with $k>2.34 h \mbox{Mpc}^{-1}$ are cut out.

\subsection{Covariance matrix estimator}
\label{subsec:covariancematrix}

The complete  description of the uncertainty about $P(k)$ is contained in a covariance matrix, defined as :
\begin{eqnarray}
C(k,k')  = \langle \Delta P(k) \Delta P(k')\rangle
\label{eq:cov_th_new}
\end{eqnarray}
where $\Delta P(k)$ refers to the fluctuations of the power spectrum about the mean.
If one has access to a large number of realizations, this matrix can be computed straightforwardly, 
and convergence can be assessed with bootstrap resampling of the realizations.
In actual surveys, however, only a handful of sky patches are observed, and a full covariance matrix extracted from these
is expected to be singular.

\begin{figure*}
  \begin{center}
  \epsfig{file=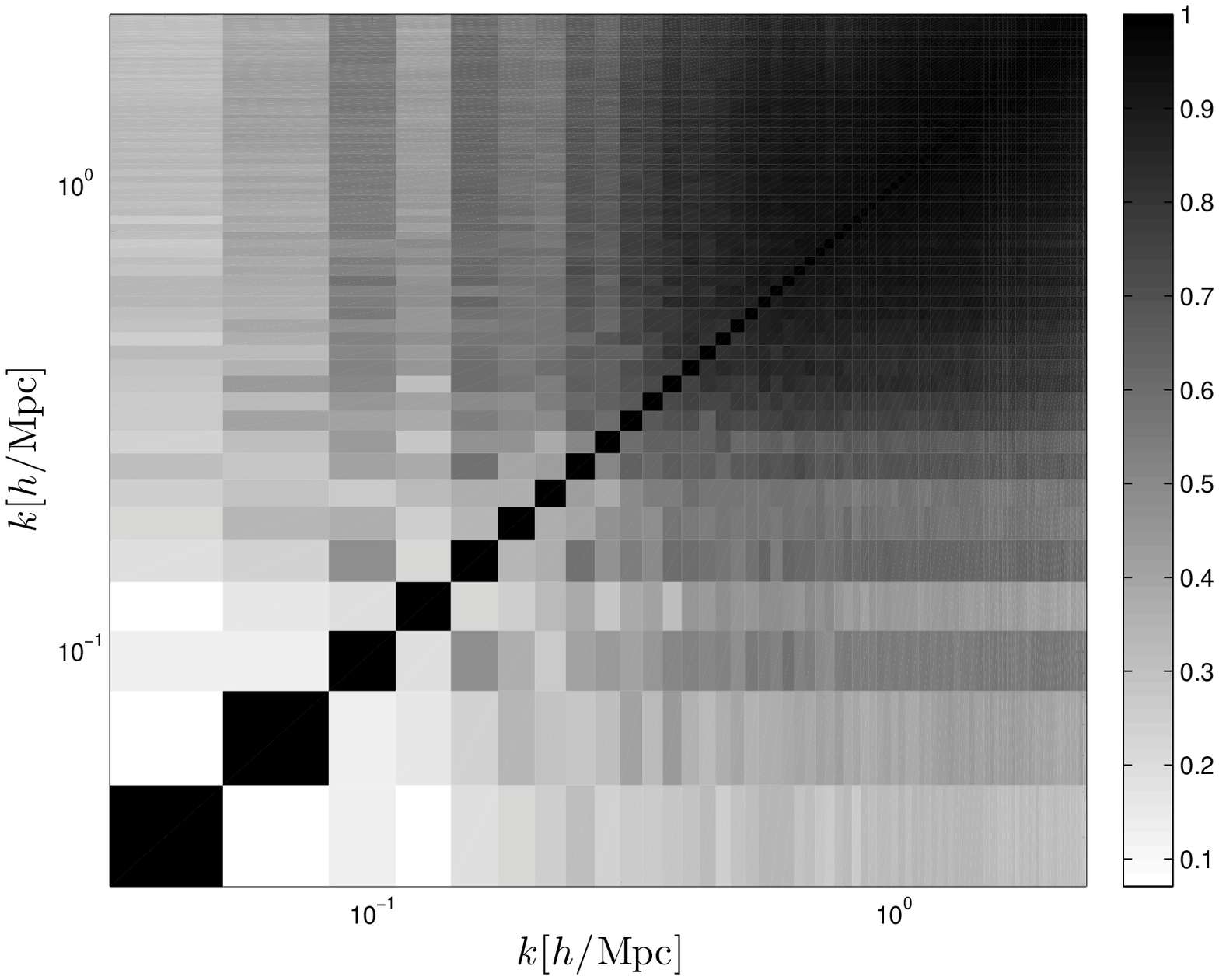,width=0.49\textwidth,height=0.49\textwidth}
  \epsfig{file=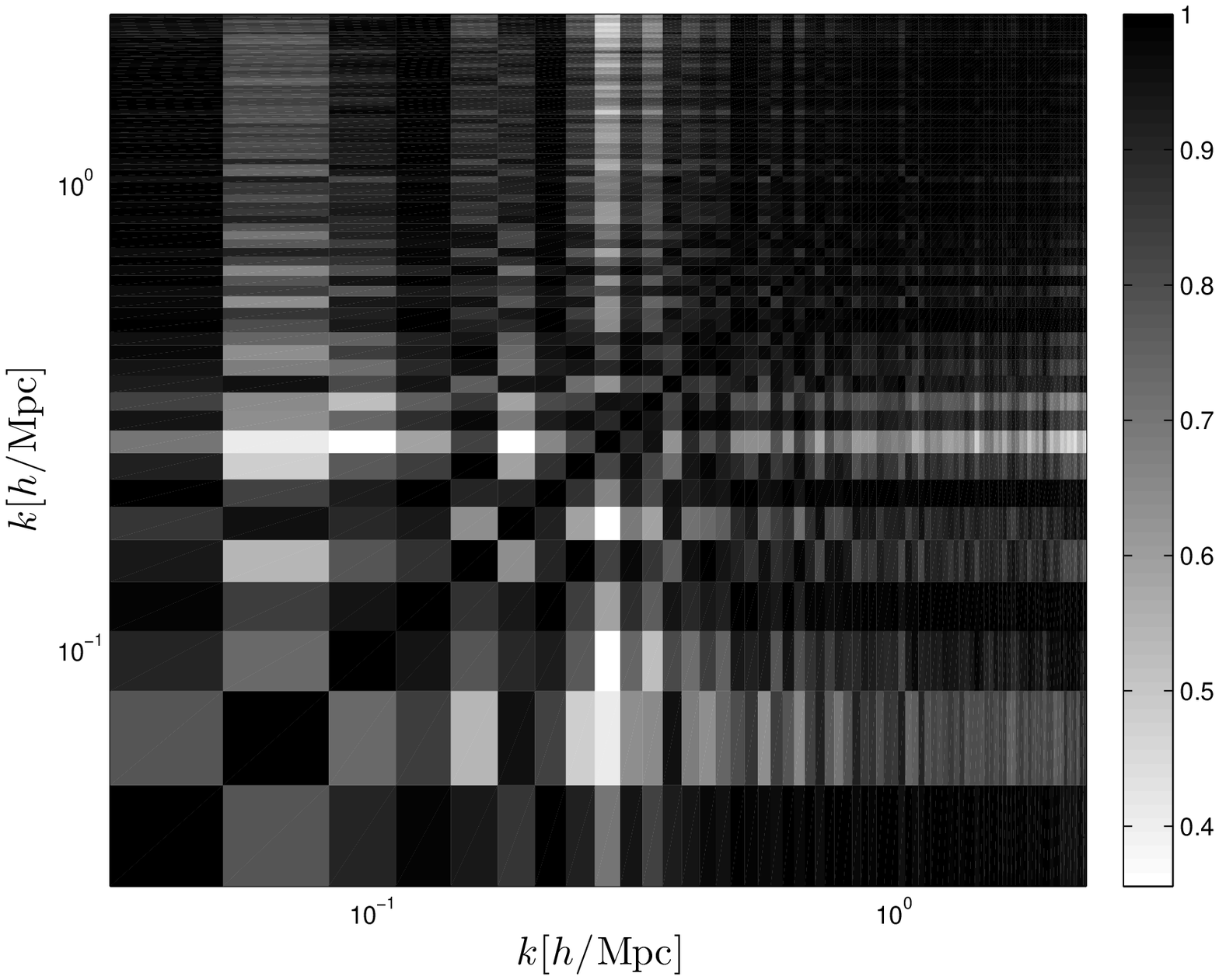,width=0.49\textwidth,height=0.49\textwidth}
    \epsfig{file=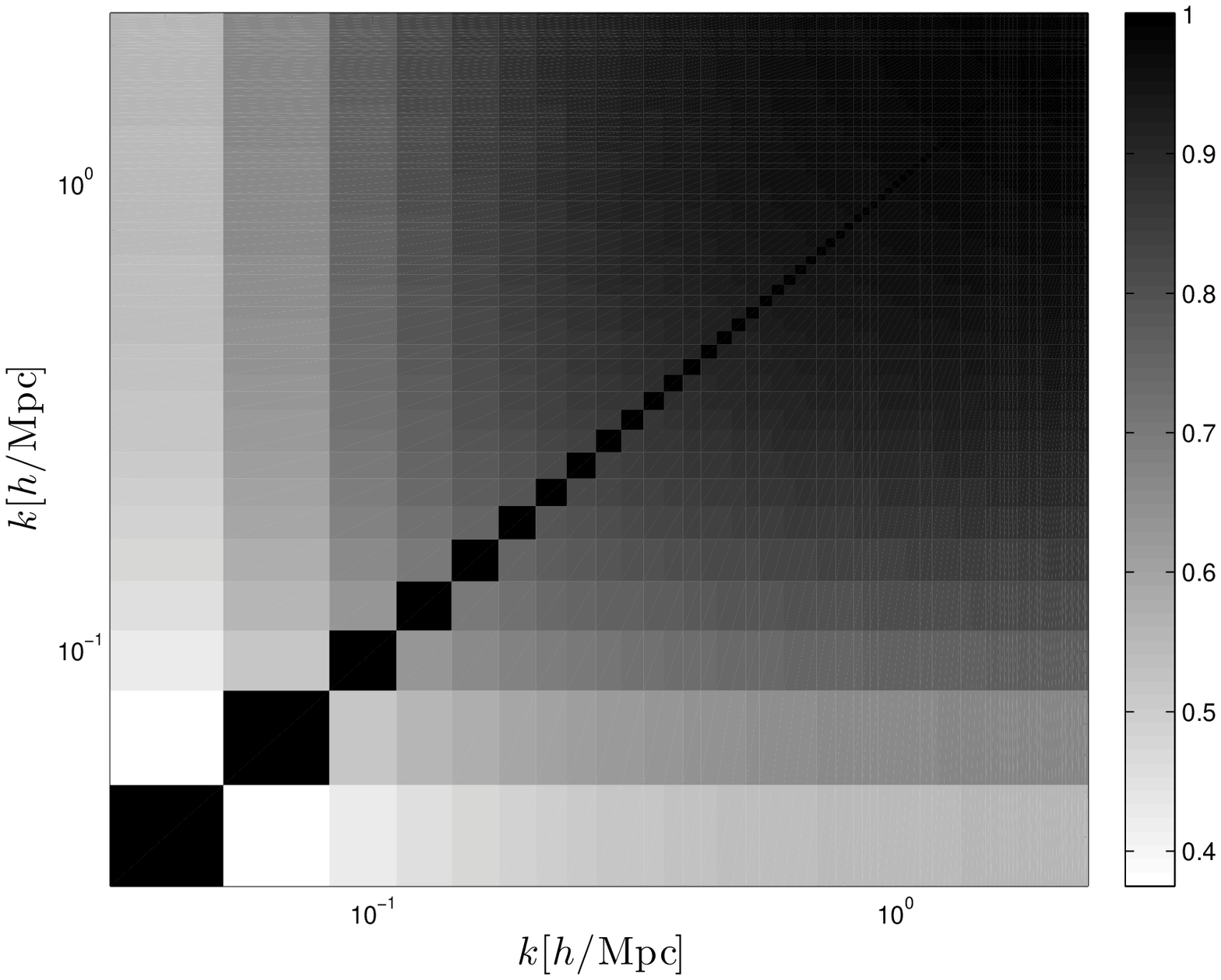,width=0.49\textwidth,height=0.49\textwidth}
    \epsfig{file=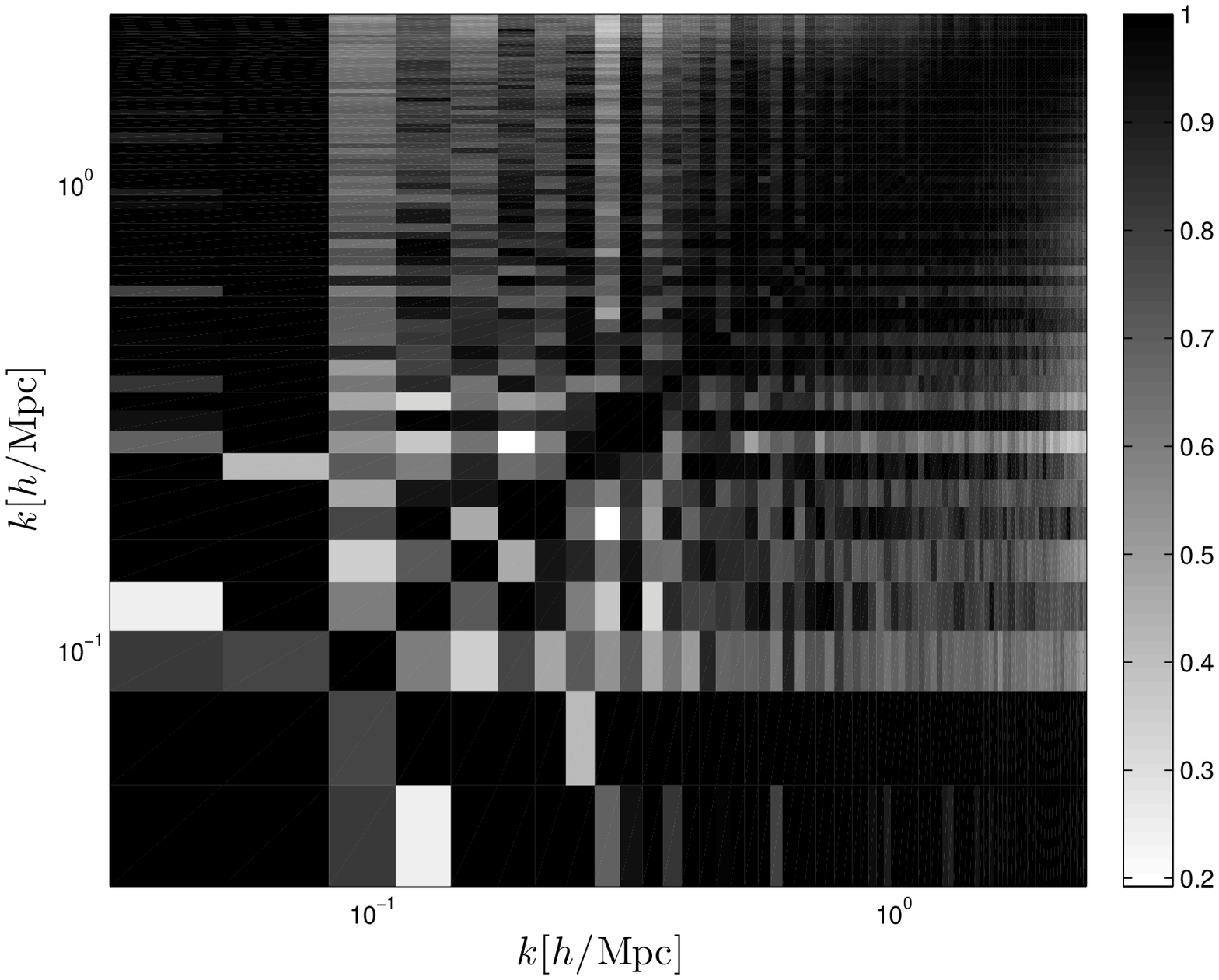,width=0.49\textwidth,height=0.49\textwidth}
  \caption{({\it top left}:) Power spectrum cross-correlation coefficient matrix, estimated from $200$ N-body simulations. 
  As first measured by \protect \citet{2005MNRAS.360L..82R}, there is a large region where the scales are correlated by 60-70 per cent. 
  This is caused by the mode coupling that occurs in the non-linear regime of gravitational systems. 
%
({\it top right}:) Same matrix as top left panel, but estimated from only $4$ simulations. 
  As expected from such a low number of measurements, the underlying structure of the matrix is partly masked by noise,
  especially in the low $k$-modes where measurements are extracted from fewer Fourier modes. Note the difference in colour scale
  between the two panels.
  ({\it bottom left}:)  Cross-correlation coefficient matrix estimated from a fit to the principal Eigenvector (the so-called `naive' way)
  of the matrix in the top right panel. 
  No other noise reduction techniques are used in this particular calculation.
  This method alone removes a lot of the noisy features in the elements,
  but yields a positive bias in the large scales (lower-left region) compared to the large sample.
  Fortunately, this region contains very few Fourier modes, hence has a minor impact on the Fisher information.  
    ({\it bottom right}:)
  Cross-correlation matrix after the Wiener filter has been applied on the $N=4$ matrix.
    There is still a significant amount of noise, and it is hard to see the improvement by eye.}
    \label{fig:cov_N4}
  \end{center}
\end{figure*}

To illustrate this, we present in the top panels of Fig. \ref{fig:cov_N4} the covariance matrices, normalized to unity on the diagonal, estimated from the $N=200$ and $N=4$ samples. 
While the former is overall smooth, we see that the latter is, in comparison, very noisy, especially in the large scales (low-$k$) regions.
This is not surprising since these large scales are measured from very few Fourier modes, hence intrinsically
have a much larger sample variance.  
It is clear why performing non-Gaussian analyses from the data is a challenging task:
the covariance matrix is singular, plus the error bar about each element is large. 
This is nevertheless what we attempt to do in this paper, and in order to overcome the singular nature of the matrix, 
we need to approach the measurement from a slightly different angle.

It was shown in HDP1 that there is an alternative way to estimate this matrix, which first requires a
measurement of $C(k,k', \theta)$, where $\theta$ is the angle between the pair of Fourier modes $({\bf k} , {\bf k'})$. 
We summarize here the properties of $C(k,k',\theta)$, and refer the reader to HDP1 for more details: 
\begin{itemize}
\item{This four-point function receives a contribution from two parts: 
the degenerate singular configuration with all ${\bf k}$ vectors
equal -- the zero-lag point -- and the smooth non-Gaussian component.}
\item{The zero-lag point corresponds to the Gaussian contribution and needs to be treated separately from the other points whenever $k=k'$.}
\item{As the angle approaches zero, the non-Gaussian component of $C(k,k', \theta)$ increases significantly, especially when both scales are in the non-linear regime.}
\item{In the linear regime, most of the contribution comes from the zero-lag point, as expected from Gaussian statistics.}
\end{itemize} 

As discussed in section $5.3$ and in the Appendix of HDP1, $C(k,k')$ can be obtained from an integration over the angular dependence of $C(k,k',\theta)$:
\begin{eqnarray}
C(k,k') & = &  \int_{-1}^1 C(k,k',\mu) d \mu   \nonumber \\
             & = & \mbox{d}\mu C(k,k',\mu=1) \delta_{kk'} + \nonumber\\
             & & \sum_{\mu   < 1} C(k,k', \mu) \mbox{d}\mu  \delta_{kk'}   +  \sum_{\mu} C(k,k', \mu) \mbox{d}\mu  (1- \delta_{kk'})  \nonumber\\
             & \equiv & G(k,k') + NG(k,k') 
\label{eq:integrate}
\end{eqnarray}
where $G$ is the first term of the second line, and $NG$ is the remaining two terms.
In the above expressions, $C(k,k',\mu=1)$ is the zero-lag point, with $\mu = \mbox{cos}\theta$.
The second line is simply in the above equation obtained by turning the integral into a sum, and splitting apart the contribution from the 
zero-lag term in the case where $k'=k$. For instance, in the case where $k' \ne k$, only the third term contributes
to $NG$, which now includes the $\theta=0$ point\footnote{Note that $NG(k,k')$ refers to the `Non-Gaussian' term, and should not be mis-read as $N \times G$. }.
When comparing  [Eq. \ref{eq:cov_th_new}]  and [Eq. \ref{eq:integrate}]  numerically, slightly different residual noise create per cent level differences,
 at least in the trans-linear and non-linear regimes\footnote{
The agreement in the linear regime is reduced by discretization effects that cause the angles to be poorly determined,
however this range is exactly where the theory is well understood. As it will become clear later in this section, the overall strategy thus puts priors on the non-Gaussian features,
requiring simply that these vanish for linear scales. }.

\subsection{$k = k'$ case}
\label{subsec:diag}

\begin{figure}
  \begin{center}
  \epsfig{file=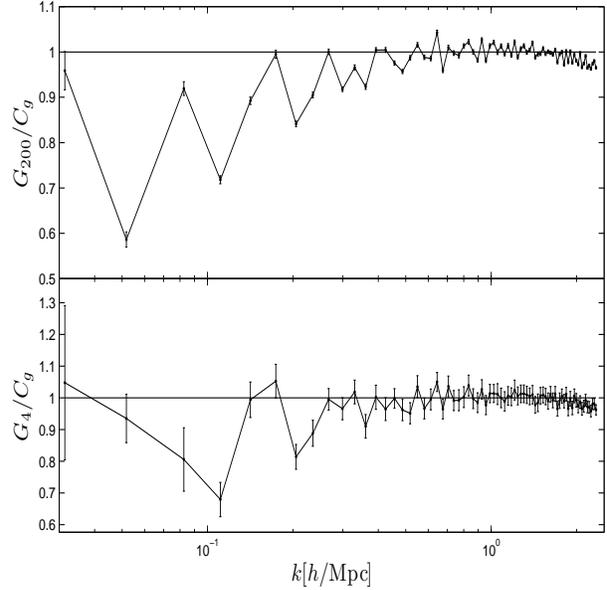,width=0.45\textwidth,height=0.45\textwidth}
  \caption{ Comparison of the Gaussian term with the analytical expression.
  The top panel shows the ratio for the $N=200$ sample, i.e. $G_{200}(k) / C_{g}(k)$, 
  while the bottom panel shows the $N=4$ ratio.
  In both cases, error bars are from bootstrap resampling. As explained in the text, the bias at low-$k$
  comes from from poor estimates of $d\mu$ in the linear regime.}
    \label{fig:zero_lag}
  \end{center}
\end{figure}

The break down of the covariance matrix proposed in  [Eq. \ref{eq:integrate}]  opens up the possibility 
to explore which of the two terms is the easiest to measure, which is noisier, 
and eventually organize the measurement such as to take full advantage of these properties.
The first term on the right hand side, $G(k,k')$, corresponds at the few per cent level -- at least in the trans-linear and non-linear regimes -- 
to the Gaussian contribution $C_{g}(k) = 2P^{2}(k)/N(k)$, 
where $N(k)$ is the number of Fourier modes in the $k$-shell.
This comparison is shown  in Fig. \ref{fig:zero_lag}, where the error bars are estimated from bootstrap resampling.
In the $N=200$ case, we randomly pick 200 realizations 500 times, and calculate the standard deviation across the measurements.
In the $N=4$ case, we randomly select 4 realizations 500 times, always from the $N=200$ sample.
In the low-$k$ regime, the simulations seem to underestimate the Gaussian predictions by up to 40 per cent.
This is not too surprising since the discretization effect is large there, and the angle between grid cell, $d\mu$, are less accurate. 
However, it appears that a substitution $G(k,k') \rightarrow C_{g}$ would improve the results by correcting for this low-$k$ bias.


\begin{figure}
  \begin{center}
   \epsfig{file=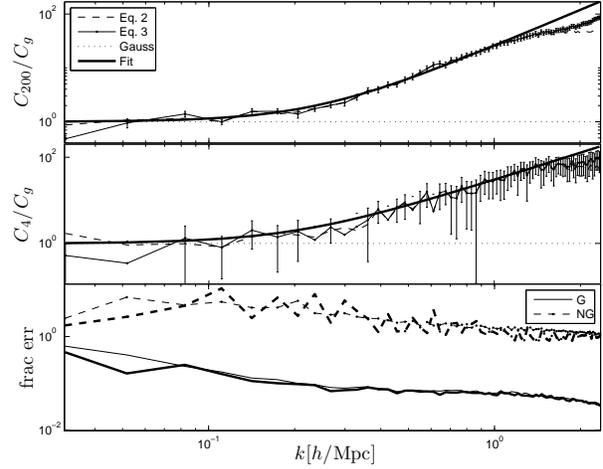,width=0.45\textwidth}
  \caption{ ({\it top}:) Ratio between the diagonal of the covariance matrix and the Gaussian analytical expression, i.e $\big(G(k) + NG(k,k)\big)/C_{g}(k)$ in the
  $N=200$ sample.
        The measurements from [Eq. \ref{eq:integrate}] are represented with the thin line, the thick solid line is from the fit, the dashed line
    is obtained from [Eq. \ref{eq:cov_th_new}], and the horizontal dotted line is the linear Gaussian prediction. 
    The error bars are from bootstrap resampling. 
     ({\it middle}:) Same as top panel, but for the $N=4$ sample. 
     This figure indicates that the confidence region for our choice of parameterization of the non-Gaussian features stops at $k=1.0 h\mbox{Mpc}^{-1}$. 
%
%
  ({\it bottom}:) Fractional error on $G$ and $NG$, scaled by $\sqrt{N}$ for comparison purposes. 
  The  thin and thick lines correspond to the $N=200$ and $N=4$ sample respectively,
  and the calculations are from bootstrap resampling. 
  It is obvious from this figure that the Gaussian term is much easier to measure than $NG$.}
    \label{fig:frac_err_GvsNG}
  \end{center}
\end{figure}

We next look at the interplay between $G$ and $NG$  on the diagonal of the covariance matrix (for the case where $k = k'$).
We know from Fig. \ref{fig:zero_lag} that the Gaussian term is  well measured and has a rapid convergence
about $C_{g}(k)$.
The top and middle panels of Fig. \ref{fig:frac_err_GvsNG} present the diagonal components of the covariance matrix, divided by the Gaussian prediction, i.e. $(G(k) + NG(k,k))/C_{g}(k)$,
for the $N=200$ and $N=4$ samples  respectively. 
The linear regime agrees well with the Gaussian statistics, then we observe strong deviations about unity as the scales become smaller.
In this figure, the error is again estimated from bootstrap resampling, even though the $G + NG$ break down allows for more sophisticated error estimates (see section \ref{sec:noiseweight}).

An important observation is that the shape of the ratio is similar for both the large and small samples, which leads us to the conclusions that 1- departures from Gaussianities 
are clearly seen even in only four fields, and 2- both samples can be parameterized 
the same way.
Following HDP1, we express the diagonal part  of $NG(k,k')$  and $C(k,k')$ as\footnote{
The notation is slightly different than
in HDP1, which expressed $C(k,k) = \frac{2P^{2}(k)}{N(k)} \bigg( 1 +  \bigg(\frac{k}{\alpha}\bigg)^{\beta} \bigg) \equiv  \frac{2P^{2}(k)}{N(k)}V(k)$. 
Note the correspondance $(\alpha,\beta) \rightarrow (k_0,\alpha)$.}:
\begin{eqnarray}
NG(k,k)  = \frac{2P^{2}(k)}{N(k)} \bigg( \frac{k}{k_{0}} \bigg)^{\alpha} \mbox{ and } 
    C(k,k) = \frac{2P^{2}(k)}{N(k)} \bigg( 1 +  \bigg( \frac{k}{k_{0}} \bigg)^{\alpha} \bigg)
\label{eq:ng_diag}
\end{eqnarray}
In this parameterization, $k_{0}$ informs us about the scale at which non-Gaussian departure become important,
and $\alpha$ is related to the strength of the departure as we progress deeper in the non-linear regime. 
The best-fitting parameters are presented in Table \ref{table:Parameters}.
As seen from Fig. \ref{fig:frac_err_GvsNG}, this simple power law form seems to model the ratio up to $k\sim 1.0 h \mbox{Mpc}^{-1}$,
beyond which the signal drops under the fit. Without a thorough check with higher resolution simulations, it is not clear 
whether this shortfall is a physical or a resolution effect. 
In the former case, we could modify the fitting formula to include the flattening observed at $k \sim 1.0 h \mbox{Mpc}^{-1}$,
but this would require extra parameters, the number of which we are trying to minimize.
We thus opt for the simple, conservative approach that consists in tightening our confidence region
and exclude modes with   $k >1.0 h \mbox{Mpc}^{-1}$, even though the simulations resolve smaller scales.

The two sets of parameters are consistent within $1\sigma$, 
which means that the $N=4$ sample has enough information
to extract the pair ($\alpha$, $k_{0}$), and therefore attempt non-Gaussian estimates of $C(k,k')$.
The fractional error on both parameters is of the order of a few per cent in the large sample,
and about 20-50 per cent in the small sample.
A second important observation is that the fractional error about $\alpha$ is about twice smaller than that of $k_{0}$,
which means that $\alpha$ is the easiest non-Gaussian parameter to extract.

\begin{table*}
\begin{center}
\caption{Best-fitting parameters of the diagonal component of the covariance matrix, 
 parameterized as $C(k,k) = G(k) + NG(k,k) = \frac{2P^{2}(k)}{N(k)} \bigg( 1 + \bigg( \frac{k}{k_{0}} \bigg)^{\alpha} \bigg)$,
 and for the principal Eigenvector, parameterized as $U(k) = A \bigg( \frac{k_{1}}{k} + 1 \bigg)^{-1}$.
The error bars are obtained from bootstrap resampling and refitting the measurements.
The Wiener filter and T-rotation techniques are described in sections \ref{subsec:wiener} and \ref{subsec:noiseweightEigen} respectively.
The parameters ($\alpha$, $k_{0}$) do not apply to the T-rotation, which acts only on the Eigenvector.}
\begin{tabular}{|l|c|c|c|c|c|}
\hline 
                                  & $k_{0}$             & $\alpha$              & $A$                           &$k_1$ \\
                  \hline
$N=200$ sample   & $0.24\pm 0.01$& $2.24\pm 0.06$ & $1.027 \pm 0.072$  & $0.07 \pm 0.02$ \\
$N=4$  sample       & $0.20\pm 0.09$& $2.1\pm 0.5$ & $1.00 \pm 0.03$      & $0.006\pm 0.02$\\
Wiener filtered         & $0.19$             & $2.01$              &  $1.00 $                     & $0.024$   \\
T-rotation                   & n/a                       & n/a                  &  $1.13 \pm 0.07$  & $0.09 \pm 0.13$\\
$G(k) = 2P(k)/N(k)$ & $0.18$             & $2.13$             &  $1.01$                   & $0.021$  \\ 

\hline 
\end{tabular}
\label{table:Parameters}
\end{center}
\end{table*}

We are now in a position to ask which of $G(k,k')$ or $NG(k,k')$ has the largest contribution to the error on $C(k,k)$.
We present in the bottom panel of Fig. \ref{fig:frac_err_GvsNG} the fractional error on both terms, in both samples.
We scale the bootstrap error by $\sqrt{N}$ to show the sampling error on individual measurements, 
and observe that in both cases, the non-Gaussian term dominates the error budget by more than an order of magnitude.

\subsection{$k \ne k'$ case}
\label{subsec:offdiag}

We now turn our attention to the off-diagonal part of the covariance matrix, whose sole contribution comes from the non-Gaussian term.
For this reason, and because there are many more elements to measure from the same data ($N^2$ vs. $N$), it is expected to be much noisier that the diagonal part. 

%
%
It is exactly this noise that makes $C(k,k')$ singular, but luckily we can filter out a large part of it 
from a principal component analysis based on the Eigenvector
decomposition of the cross-correlation coefficient matrix $r(k,k') \equiv C(k,k')/\sqrt{C(k,k)C(k',k')}$.
As discussed in \citet{2012MNRAS.419.2949N}, this method improves the accuracy of both the covariance matrix and of its inverse.
HDP1 further provides a fitting function for the Eigenvector, 
and we explore here how well the best-fitting parameters can be found in a low statistics environment. 

\begin{figure*}
  \begin{center}
  \epsfig{file=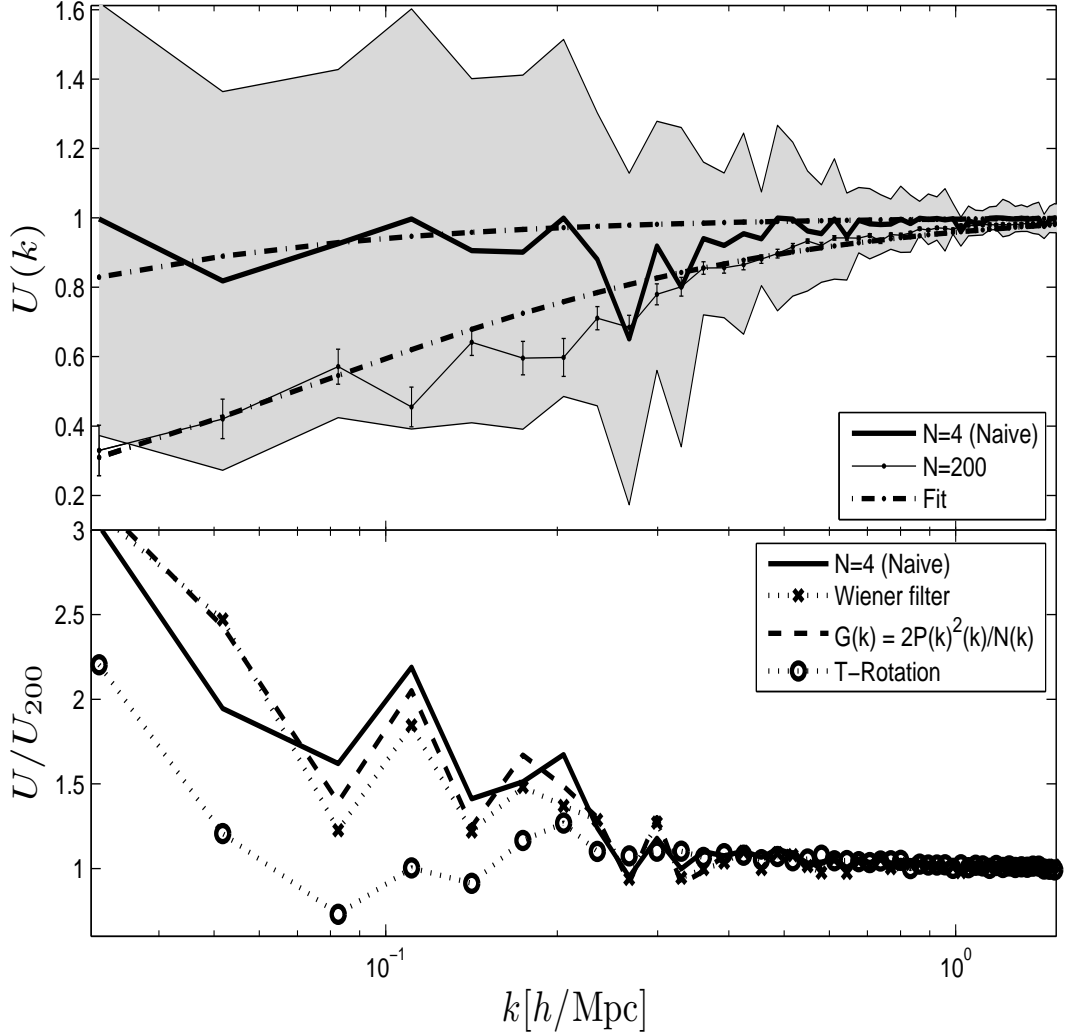,width=0.80\textwidth,height=0.80\textwidth}
  \caption{Main Eigenvector extracted from the different estimates of the cross-correlation coefficient matrix. 
  As explained in the text, the Eigenvalue is absorbed in the vector to simplify the comparison.
  ({\it top}:) The thin solid line (with error bars) and the thick solid line (with grey shades) represent the $N=200$ and $N=4$ measurements respectively.
  Since no other noise reduction technique are applied to extract these vectors, we refer to this $N=4$ estimate as the `naive' estimate. 
   The error bars are from bootstrap, and the fits to these curves are provided with the parameters of Table \ref{table:Parameters} and represented
   by the dot-dashed lines. 
  ({\it bottom}:) Ratio between the main Eigenvector extracted from three noise reduction techniques, and that of the $N=200$ sample 
 ( shown in the upper panel).  The $G(k) \rightarrow 2P^2(k)/N(k)$ substitution is described in section \ref{subsec:diag},
 while the  Wiener filter and T-rotation approaches are described in sections \ref{subsec:wiener} and \ref{subsec:noiseweightEigen} respectively.
 Except for the first bin, the T-rotation technique recovers the $N=200$ Eigenvector to within $20$ per cent for $k<0.2 h \mbox{Mpc}^{-1}$;
 all techniques achieve per cent level precision for smaller scales, due to the higher number of modes per $k$-shell.   
    }
    \label{fig:U}
  \end{center}
\end{figure*}

This decomposition is an iterative process that factorizes the cross-correlation coefficient matrix into
a purely diagonal component and a smooth  symmetric off-diagonal part.
The latter is further Eigen-decomposed,
and we keep only the Eigenvector $U(k)$ that corresponds to the largest Eigenvalue $\lambda$. 
In that case, it is convenient to absorb the Eigenvalue in the definition of the Eigenvector, i.e. $\sqrt{\lambda} U(k) \rightarrow U(k)$, 
such that we can write the $r(k,k') \sim U(k)U(k')$ directly. 
Since the diagonal elements are unity by construction, the exact expression is $r(k,k') =   U(k)U(k') + \delta_{kk'} [1 - U^{2}(k)]$.
This effectively puts a prior on the shape of the covariance matrix, since any part of the signal that does not fit this shape 
is considered as noise and excluded.
As shown in HDP1, this decomposition is accurate at the few per cent level in the dynamical range of interest,
for the $N=200$ sample.
We present in the top panel of Fig. \ref{fig:U} the Eigenvector extracted from both samples, against a fitting function of the form\footnote{This parameterization
of $U(k)$ is equivalent to that of HDP1, but significantly simplified: we factorize $\gamma$ outside of the parenthesis, and we set $\delta = 1$.
This choice is motivated by the fact that in the bootstrap estimate, the error bar on $\delta$ was at the fraction of a per cent,
indicating a weak dependence of the fit on this variable.}:
\begin{eqnarray}
U(k) = \frac{A}{k_{1}/k + 1}
\label{eq:fit_U}
\end{eqnarray}
In this parameterization, $A$ represents the overall strength of the non-Gaussian features, while $k_{1}$ is related to the scale
where cross-correlation becomes significant in our measurement.
If $A=0$, then we recover $r(k,k') = \delta_{kk'}$, which corresponds to `no mode cross-correlation'.
The best fitting parameters are shown in Table \ref{table:Parameters}.
We observe that the error bar on $A$ is at the few per cent level, and that the measurements in both samples are both very close to unity. 
On the other hand, $k_{1}$ is much harder to constrain: in the $N=200$ sample, the fractional error bar 
is about $30$ per cent, and in the $N=4$, the error bar is three times larger than its mean. 
In other words, $k_{1}$  is consistent with zero within $1\sigma$ in the small sample, which corresponds to a constant Eigenvector.
The non-Gaussian noise is thus mostly concentrated here, and improving the measurement on $k_{1}$ is one of the main
tasks of this paper.

%

If we model the $N=4$ covariance matrix with this fit to the Eigenvector, the matrix is no longer singular, as shown in the bottom left panel of Fig. \ref{fig:cov_N4}.
It does exhibit a stronger correlation at the largest scales, compared to the matrix estimated from the large sample;
this difference roots in the fact that the $N=4$ Eigenvector remains high at the largest scales, whereas the $N=200$ vector drops.
There is thus an overestimate of the amount of correlation between the largest scales, which biases the uncertainty estimate on the high side.
In any case, these large scales are given such a low weight  in the calculation of Fisher information -- they
contain a small number of independent Fourier modes -- that their contribution to the Fisher information is tiny,
as will become clear in section \ref{sec:fisher}. 
The bottom left panel of Fig. \ref{fig:cov_N4} also shows that we do recover the region where the cross-correlation coefficient is 60-70 per cent,
also seen in the top left panel of the figure. It is thus a significant step forward in the accuracy of the error estimate compared to the Gaussian approach.

We show in sections \ref{subsec:wiener} and \ref{subsec:noiseweightEigen} that noise filtering techniques are able to reduce this bias down to a minor effect, 
even when working exclusively with the same four fields. To contrast the pipeline presented in this section (of the form  $[N=4 \rightarrow \mbox{ PCA } \rightarrow \mbox{ fit }]$) 
to those presented in future sections,  we hereafter refer to this approach as the `$N=4$ naive' way. 



\section{Optimal Estimation of $C(k,k')$: Beyond bootstrap}
\label{sec:noiseweight}

In the calculations of section \ref{sec:bkg}, we show that even in low statistics, 
we can extract four non-Gaussian parameters with the help of noise filtering techniques that
assume a minimal number of priors on the non-Gaussian features.
As mentioned therein, all the error bars are obtained from bootstrap resampling,
which is generally thought to be a faithful representation of the underlying variance only in the large sample limit.
How, then, can we trust the significance of our results in the $N=4$ sample, which emulates an actual galaxy survey?
More importantly, can we do better than bootstrap?
In this section, we expose new procedures that optimize the estimate of both the mean and the error {\it on} $C(k,k')$,
and we quantify the improvements on all four non-Gaussian parameters ($\alpha$, $A$, $k_0$ and $k_1$).


\subsection{Wiener filtering}
\label{subsec:wiener}

Let us recall that in the bootstrap estimate of the error on $C(k,k')$,
we resample the measured $C(k,k',\theta)$, integrate over the angle $\theta$ and add up the uncertainty from different angles -- 
including the zero-lag point -- in quadrature.
We propose here a different approach, based on our knowledge that $C(k,k',\theta)$ is larger for smaller angles.
At the same time, we take advantage of the fact that the noise on $G(k)$ is much smaller than that on $NG(k,k')$.
Since the mean and error should be given more weight in regions where the signal is cleaner,
we replace the quadrature by a noise-weighted sum in the angular integration of $C(k,k',\theta)$.
This replacement reduces the error on $NG$ by an order of magnitude or so, depending on the scale 
(see Appendix \ref{app:noise_weighted} for details). 


%
%


At this stage, we now have accurate estimates of the {\it error on} the covariance $C(k,k')$ in the
$N=4$ sample, as well as accurate measurements of the signal and noise of the underlying covariance matrix from the $N=200$ sample, which we treat as a {\it template}. 
The technique we describe here is a Wiener filtering approach that uses known noise properties of the system in order to extract a signal that is closer to the template.
The error on $G(k)$ is obtained from bootstrap resampling the zero-lag point,
while the error on $NG(k,k')$ comes from the noise-weighted approach mentioned above and discussed in Appendix
\ref{app:noise_weighted}. We first apply the filter on both quantities separately, and then combine the results afterward.
Namely, we define our Wiener filters as
\begin{eqnarray}
C_{WF} = G_{WF} + NG_{WF}
\label{eq:wienersum}
\end{eqnarray}
with 
\begin{eqnarray}
G_{WF} = G_{200} + (G_{4} - G_{200})\left(\frac{\sigma^{2}_{200}}{\sigma^{2}_{200} - \sigma^{2}_{4}}\right)
\label{eq:wienerfilter}
\end{eqnarray}
and
\begin{eqnarray}
NG_{WF} = NG_{200} + (NG_{4} - NG_{200})\left(\frac{\sigma^{2}_{200}}{\sigma^{2}_{200} - \sigma^{2}_{4}}\right)
\label{eq:wienerfilter2}
\end{eqnarray}
Note that the errors that appear in the above two expressions correspond to the estimates from the $N=200$ and $N=4$ samples  on $G$ and $NG$ respectively.
 
\begin{figure}
  \begin{center}
    \epsfig{file=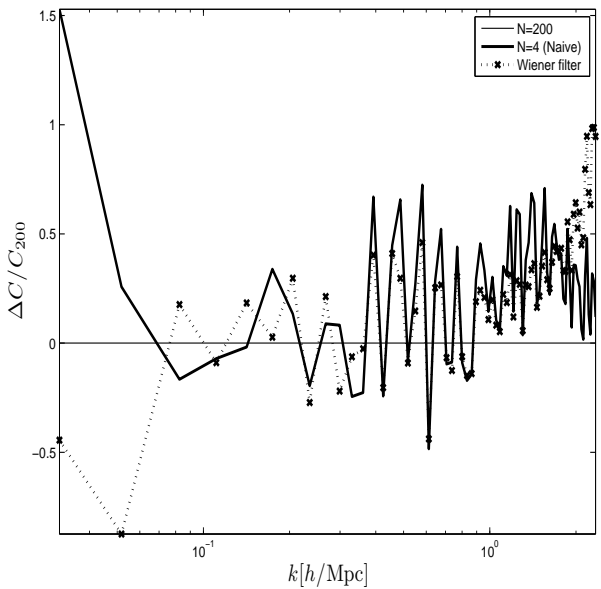,width=0.45\textwidth,height=0.45\textwidth}
  \caption{ Comparison between the variance about $P(k)$ with and without 
  the Wiener filtering technique. Results are expressed as the fractional error with respect to the $N=200$ sample,
  acting as a signal template in our filter.  This plot shows that Wiener filtering can reduce the noise 
  on the diagonal elements, as seen by the smaller fluctuations about the $N=200$ sample.}
    \label{fig:cov_diag_N4_WF}
  \end{center}
\end{figure}

 We present in Fig. \ref{fig:cov_diag_N4_WF} the  Wiener filtered variance on $P(k)$, compared to the $N=200$ and $N=4$ samples.
We observe in the range $0.3 < k < 2.0 h\mbox{Mpc}^{-1}$ that the filter decreases the size of the fluctuations about the $N=200$ sample, compared to the original $N=4$ sample.
For larger scales, it is not clear that the effect of the filter represents a gain in accuracy: 
while the variance on the fundamental mode is 3 times closer to the template's measurement, the variance about the second largest mode does 3 times worst, 
while the change in others large modes seems to have no gain.
However, we recall that very little weight is given to these low-$k$ modes in the calculation of the Fisher information about the dark matter power spectrum.
Hence slightly degrading the accuracy of the variance about the two largest modes is a mild cost if we can improve the range that ultimately matters, i.e. $0.1 < k < 1.0 h\mbox{Mpc}^{-1}$.
In addition, our parameterization of the non-Gaussian features assign a smaller weight to the large scales, which are smoothly forced towards 
the analytical Gaussian predictions.

There is a positive bias of about 15 per cent and up that appears for $k>1.0 h\mbox{Mpc}^{-1}$, both in the original $N=4$ sample and in the Wiener filtered product. 
The bias in the unfiltered fields is simply a statistical fluctuation, since we know it does converge to the large sample by increasing the number of fields.
It does, however,  propagate in a rather complicated way through the Wiener filter, causing a sharp increase at $k>1.0 h\mbox{Mpc}^{-1}$.
Because the details of how and why this happens are rather unclear, we decided to exclude this region from the analysis, based on suspicion of 
systematics. 
The full Wiener filtered cross-correlation matrix is presented in the bottom right panel of Fig. \ref{fig:cov_N4} and shows that some of the noise has been filtered out:
the regime  $k >0.5 h\mbox{Mpc}^{-1}$ is  smoother than the original $N=4$ measurement, 
and except for the largest two modes, the matrix is brought closer to the $N=200$ sample. 

The next steps consist in computing the new Eigenvector $U(k)$ constructed with $C_{WF}$
and to find the new best-fitting parameters  ($\alpha$, $k_{0}$, $A$, $k_{1}$).
The bottom panel of Fig. \ref{fig:U} presents $U(k)$ and shows that most of the benefits are seen for $0.08 < k < 0.2 h\mbox{Mpc}^{-1}$, which is the range we targeted to start with,
and the best-fitting parameters are tabulated in Table \ref{table:Parameters}.
Overall, the improvement provided by the Wiener filter is still hard to gauge by eye from Fig. \ref{fig:cov_N4}, 
because the Eigenvector is still very noisy. This is to be expected: the method is mostly efficient on the diagonal part,
where we can take advantage of the low noise level of the Gaussian term. 


\subsection{Noise-weighted Eigenvector decomposition}
\label{subsec:noiseweightEigen}


In this section, we describe a last noise filtering technique that utilizes
known properties of the noise about the Eigenvectors and their Eigenvalues to improve the way we perform the 
principal component decomposition in the $N=4$ sample.
It is a general strategy that could be combined with others techniques described in the preceding sections,
however, for the sake of clarity, we only present here the standalone effect.  

\begin{figure}
  \begin{center}
    \epsfig{file=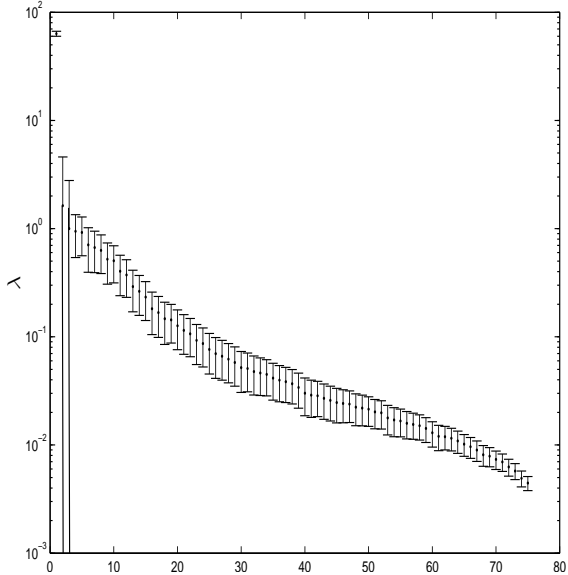,width=0.45\textwidth,height=0.45\textwidth}
  \caption{Signal and noise of the Eigenvalues measured from the $N=200$ sample. 
  The error bars are obtained from bootstrap resampling the power spectrum measurements.
  Only the four largest Eigenvalues are kept in the analyses.} 
  \label{fig:Eigenvalues}
  \end{center}
\end{figure}

In the Eigen-decomposition, not all Eigenvalues are measured with the same precision. 
For instance, most of the covariance matrix can be described by the first Eigenvector $U(k)$, 
hence we expect the signal-to-noise ratio about its associated Eigenvalue to be the largest.
Considering again the $N=200$ sample as our template of the underlying covariance, 
the error on each $\lambda$ can be obtained by bootstrap resampling the 200 realizations. 
We present this measurement in Fig. \ref{fig:Eigenvalues}, 
where we observe that the first Eigenvalue is more than an order of magnitude larger than the others, which are also noisier. 

Since general Eigenvector decompositions are independent of rotations, our strategy is to rotate the $N=4$ cross-correlation matrix
into a state ${\mathbf T}$ where it is brought closer to the template, then apply a signal-to-noise weight before 
the Eigenvector decomposition.
More precisely, we apply the following algorithm, which we refer to as the `T-rotation' method (for rotation into ${\mathbf T}$-space) in the rest of this paper: 
\begin{enumerate} 
\item{Rotate the noisy (i.e. $N=4$) cross-correlation coefficient matrix  in the Eigenstates ${\mathbf T}$ of the template}
\item{Weight the elements by the signal-to-noise ratio of the corresponding Eigenvalues} 
\item{Perform an Eigenvector decomposition on the resulting matrix}
\item{Undo the weighting}
\item{Rotate back}
\end{enumerate}

The rotation in step (i) is defined as:
\begin{eqnarray}
{\mathbf R} = {\mathbf T}^{-1} {\mathbf \rho} {\mathbf T}
\label{eq:rotation}
\end{eqnarray}
and,  by construction, reduces to the diagonal Eigenvalues matrix in the case where ${\mathbf \rho}$ is the template cross-correlation coefficient matrix.
The weighting in step (ii) is performed in two parts\footnote{The first part takes the ${\mathbf R_{template}}$ matrix into the identity matrix, and allows the second part to done with the correct units.}: 
1- we scale each matrix element $R_{ij}$ by $1/\sqrt{\lambda_{i} \lambda_{j}}$, and 
2- we weight the result by the signal to noise ratio of each $\lambda$. 
Combining, we define\footnote{This matrix is not to be confused with the fluctuation matrices of [Eq. \ref{eq:noiseweightangle}].}:   
\begin{eqnarray}
D_{ij} \equiv  R_{ij} \frac{1}{\sqrt{\lambda_{i} \lambda_{j}}}\left(\frac{\lambda_{i} \lambda_{j}}{\sigma_{i} \sigma_{j}} \right)^{2} \equiv R_{ij} w_{i} w_{j}
\label{eq:noiseweight}
\end{eqnarray}
As seen in Fig. \ref{fig:Eigenvalues}, the Eigenvalues drop rapidly, and we expect only the first few to contribute to the final result.
In fact, our results present very small variations if we keep anywhere between  two and six Eigenvalues and exclude the others. 
Since it was shown in HDP1 that in some occasions, we need up to four Eigenvectors to describe the observed $C(k,k')$ matrix,
we choose to keep four Eigenvalues as well, and cut out the contributions from $\lambda_{i>4}$. 

In step (iii) the resulting matrix ${ \mathbf D}$ is decomposed into Eigenvectors ${\mathbf S}$:
\begin{eqnarray}
{\mathbf D} = {\mathbf S} {\mathbf \lambda_D} {\mathbf S^{-1}} \mbox{\vspace{1cm}} 
\label{eqn:rotateback}
\end{eqnarray}
Then, in step (iv), these Eigenvectors are weighted back by absorbing the weights directly:
\begin{eqnarray}
\widetilde{S_{ij}} = S_{ij}/w_i
\end{eqnarray} 
The result is finally rotated back into the original space:
\begin{eqnarray}
\tilde{\mathbf \rho} = \tilde{\mathbf T} {\mathbf \lambda_D} \tilde{\mathbf T^{-1}} \mbox{\vspace{1cm} with \vspace{1cm}}  \tilde{\mathbf T} =  {\mathbf T} \tilde{\mathbf S}
\end{eqnarray} 
If no cut is applied on the Eigenvalues,  this operation essentially does nothing to the matrix, as the equivalence between $\tilde{\mathbf \rho}$ and ${\mathbf \rho}$ is exact: every rotation and weights that are applied are removed, and we get  $U(k) \equiv \tilde{T}_{1}(k)$. 
However, the cut, combined with the rotation and weighting, acts as to improve the measurement of the Eigenvector. 
Physically, this is enforcing on the measured matrix a set of priors, corresponding to the Eigenvectors of the template, 
with a strength that is weighted by the known precision about the underlying Eigenvalue. 
For a Gaussian covariance matrix normalized to 1 on the diagonal, all frames are equivalent, and any rotation from any prior has no impact on the accuracy.  
When the covariance is not diagonal, however, some frames are better than others, and the Eigenframe is among the best,  
as long as the simulation Eigenframe is similar to that of the actual data.  If these frames were unrelated, this T-rotation would generally be neutral.

We present in the bottom panel of Fig. \ref{fig:U} the effect of this T-rotation on the original $N=4$ Eigenvector.
We observe that it traces remarkably well the $N=200$ vector, to within $20$ per cent even at the low $k$-modes,
and outperforms the other techniques presented in this paper in its  extraction of  $U(k)$.
The best fitting parameters corresponding to $\tilde{T}_{1}(k)$ are summarized in Table \ref{table:Parameters}.
We discuss in section \ref{sec:recipe} how, in practice, one can us these techniques in a real survey.

\section{Impact on Fisher Information}
\label{sec:fisher}

In  analyses based on measurements of $P(k)$, the uncertainty typically
propagates to cosmological parameters within the formalism of  Fisher matrices \citep{Tegmark:1997rp}.
The Fisher information content in the amplitude of the power spectrum, defined as 
\begin{eqnarray}
I(k_{max}) = \sum_{k,k' < k_{max}}C_{norm}^{-1}(k,k')|_{k,k' < k_{max}}
\end{eqnarray}
effectively counts the number density of degrees of freedom
in a power spectrum measurement. 
In the above expression, $C_{norm}$ is simply given by $C(k,k')/[P(k)P(k')]$.
The Gaussian case is the simplest, since the covariance is given by $C_{g}(k) = 2P^2(k)/N(k)$,
where $N(k)$ is the number of cells in the $k$-shell. 
We recall that the factor of two comes in because the $P(-{\bf k}) = P({\bf k})$ symmetry,
which reduces the number of independent elements  by a factor of two. 
Also,  $C_{norm}$ reduces to $2/N(k)$, and $I(k_{max}) = \sum_{k} N(k)/2 $ for Gaussian fields.

We see how $I(k)$ is an important intermediate step to the full Fisher matrix calculation, as it 
tells whether we can expect an improvement on the Fisher information from a given increase in survey resolution.
It was first shown by \citet{2005MNRAS.360L..82R} that the number of degrees of freedom
increases in the linear regime, following closely the Gaussian prescription, but then reaches a trans-linear plateau,
followed by a second increase at even smaller scales.
This plateau was later interpreted as a transition between the two-haloes and the one-halo term \citep{Neyrinck:2006xd},
and corresponds to a regime where the {\it new} information is degenerate with that of larger scales. 
By comparison, the Gaussian estimator predicts ten times more degrees of freedom by $k \sim 0.3 h\mbox{Mpc}^{-1}$.

What stops the data analyses from performing fully non-Gaussian uncertainty calculations 
is that the Fisher information requires an accurate measurement of the {\it inverse} of a covariance matrix similar to that seen in Fig. \ref{fig:cov_N4},
which is singular. 
With the noise reduction techniques described in this paper, however, the covariance matrix is no longer singular,
such that the inversion is finally possible.
To recapitulate, these techniques are:  
\begin{enumerate}
\item{The `Naive $N=4$ way' : straight Eigenvector decomposition + fit of the $N=4$ sample (section \ref{subsec:offdiag})}
\item{Same as (i), with the $G(k) \rightarrow \frac{2 P(k)^2}{N(k)}$ substitution (section \ref{subsec:diag})}
\item{Wiener filtering of $G(k)$ and $NG(k,k')$ (section \ref{subsec:wiener})}
\item{T-rotation (section \ref{subsec:noiseweightEigen})}
\end{enumerate}
In this section, we assume that there are no survey selection function effects, and that the universe is periodic.
We discuss more realistic cases in section \ref{sec:recipe}.

\begin{figure*}
  \begin{center}
  \epsfig{file=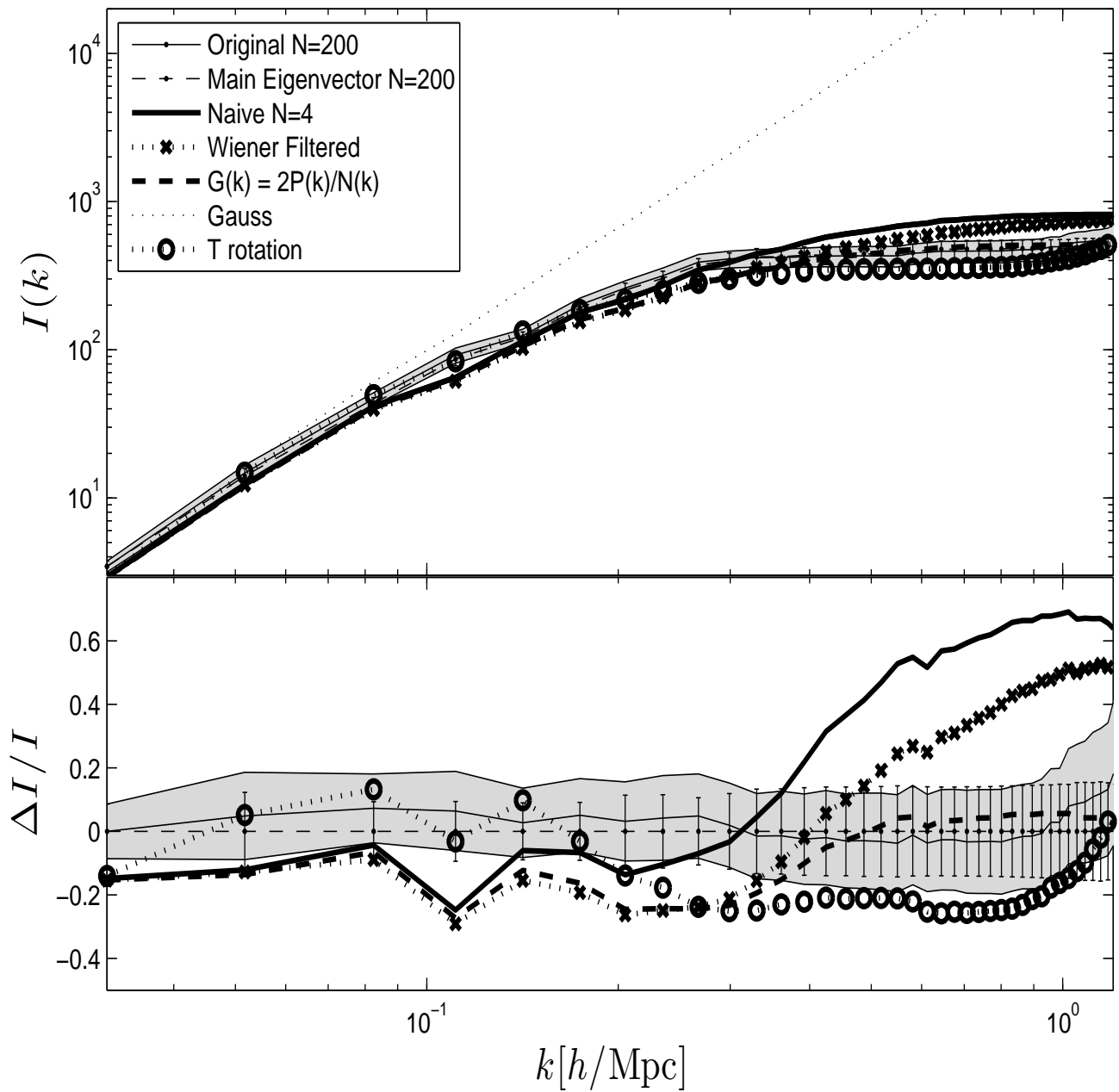,width=0.98\textwidth,height=0.98\textwidth}
  \caption{ ({\it top}:) Fisher information extracted from various noise filtering techniques, compared to Gaussian calculations. 
  Results from the original $N=200$ sample are shown by the thin solid line, with error bars plotted as grey shades; 
 calculations from the main Eigenvector of the $N=200$ matrix are shown by the dashed line, with error bars;
  results from fitting directly the main Eigenvector of the $N=4$ sample (see section \ref{subsec:offdiag}) are shown by the thick solid line;
   the effect of replacing $G(k) \rightarrow 2P(k)/N(k)$ (+ fit) is shown by the thick dashed line (see section \ref{subsec:diag});
   the effect of Wiener filtering the covariance matrix (+ fit) is shown by the crosses (see section \ref{subsec:wiener});
  finally,  the noise-weighted T-rotation calculations (+ fit) are shown by the open circles (see section \ref{subsec:noiseweightEigen}).
({\it bottom}:) Fractional error between each of the top panel curves and the measurements from the main $N=200$ Eigenvector.}
    \label{fig:Fisher}
  \end{center}
\end{figure*}

We present in the  top panel of Fig. \ref{fig:Fisher} the Fisher information content 
for each technique, compared to that of the template and the analytical Gaussian calculation.
We also show the results for the $N=200$  sample after the Eigen-decomposition,
which is our best estimator of the underlying information \citep{2012MNRAS.419.2949N}.
The agreement between this and the original information content in the $N=200$ sample is at the few per cent level for $k<1.0 h\mbox{Mpc}^{-1}$ anyway. 
We do not show the results from the $N=4$ sample, nor that after the Eigenvector decomposed only, as the curve quickly diverges. 
It actually is the fitting procedure, summarized in Table \ref{table:Parameters},  that clean up enough noise to make the inversion possible.
In all these calculations, the error bars are obtained from bootstrap resampling.
The bottom panel  represents the fractional error between the different curves and our best estimator.
 
 As first found by \citet{2005MNRAS.360L..82R}, the deviation from Gaussian calculations reaches an order of magnitude by $k \sim 0.3 h/\mbox{Mpc}$, 
 and increases even more at smaller scales.
 With the fit to the Eigenvector (technique (i) in the list above mentioned), we are able to recover a Fisher information content much closer to the template; 
 it  underestimates the template by less than $20$ per cent  for $k < 0.3 h \mbox{Mpc}^{-1}$, 
and then overestimates the template by less than  $60$ per cent away for $0.3 < k < 1.0 h \mbox{Mpc}^{-1}$.
The fit to the analytical substitution of $G(k)$ (technique (ii)) has even better performances, with maximum deviations of $20$ per cent over the whole range.
The fit to the Wiener filter (iii) is not as performant, but still improves over the naive $N=4$ way, with the maximal deviation reduced to less than $45$ per cent.
Finally, the fit to the T-rotated (iv) Eigenvectors  also performs well, with deviations by less than $20$ per cent.  
Most of the residual deviations can be traced back to the fact that in the linear regime, the covariance matrix exhibited a large noise that
 we could not completely remove. This extra correlation translates into a loss of  degrees of freedom in the linear regime, a cumulative effect that biases  the Fisher information content
 on the low side. Better noise reduction techniques that focus in the large scales cleaning could outperform the current information measurement.
In any case, this represents a significant step forward for non-Gaussian data analyses since estimates of $I(k)$ can be made accurate to within $20$ per cent 
over the whole BAO range, even from only four fields, with minimal prior assumptions.
 
\section{In the Presence of a Survey Selection Function}
\label{sec:recipe}

The results from section \ref{sec:fisher} demonstrate that it is possible to 
extract a non-Gaussian covariance matrix internally, from a handful
of observation patches, and that with noise filtering techniques, 
we can recover, to within $20$ per cent,  the Fisher information content in the amplitude of the power spectrum
of (our best estimate of) the underlying field.
The catch is that these are derived from an idealized environment that exist only in N-body simulations,
and the objective of this section is to understand how, in practice, can we
apply the techniques in actual data analyses.
\footnote{As mentioned in the introduction of this paper, the key missing ingredient in this paper
is the inclusion of redshift distortions, which will be the core subject of the third paper of this series. }
We explore a few simple cases that illustrates how the noise reduction techniques can be applied,
and how the non-Gaussian parameters can be extracted 
in the presence of a survey selection function $W({\bf x})$.


\subsection{Assuming deconvolution of $W({\bf k})$} 
\label{subsec:deconvolved}

The first case we consider is the simplest realistic scenario one can think of, in which the observation patches 
are well separated, and for each of these the survey selection functions can be successfully deconvolved from the
underlying fields. 
For simplicity, we also assume that each patch is assigned onto a cubical grid with constant volume, resolution and redshift, such that the
observations combine essentially the same way as the $N=4$ sample presented in this paper.
Once the grid is chosen, one then needs to produce a large sample of realizations from N-body simulations, 
with the same volume, redshift and accuracy, and construct the equivalent of our $N=200$ sample.
\begin{enumerate}
\item{In the `naive $N=4$' way, one needs to compute the (noisy) covariance matrix from the data sample,
compute the ratio of the diagonal to the Gaussian predictions and fit, then compute the cross-correlation coefficient matrix $\rho(k,k')$,
Eigen-decompose and fit the main Eigenvector $U(k)$. The noise filtered estimate of the covariance is recovered by combining the 
fitting functions at the centre of the $k$-bins for the ratio and $U(k)$, using $\rho(k,k') = U(k)U(k')$ and $C(k,k') = \rho(k,k') \sqrt{C(k,k) C(k',k')}$.}
\item{To construct the Wiener filter described in section \ref{subsec:wiener}, one needs to use the methodology of HDP1
and compute $C(k,k',\theta)$ from the density fields of the data and the large simulated sample,
and finally extract $G(k)$ and $NG(k,k')$ from [Eq. \ref{eq:integrate}] to construct the filter.
At that stage, it is also trivial to try the semi-analytical substitution $G(k) \rightarrow 2P^2(k)/N(k)$ and reduce the noise even more.
After that, we need to compute the ratio and $U(k)$ as described above, find the best-fitting parameters, and reconstruct the covariance matrix.}
\item{To make use of the T-rotation technique, one needs to compute, from the large simulated sample, 
the Eigenvectors that describe the cross-correlation coefficient matrix, plus the noise about each Eigenvalue, 
which can be obtained from bootstrap resampling the simulated realizations.
The weights $w_i$ can then be computed, and the rest of the technique follows directly from section \ref{subsec:noiseweightEigen},
such that we end up with a better estimate of $U(k)$ -- to be fitted as well.
This technique improves only the estimate of $\rho$, 
so that one then has some freedom regarding which estimate of the diagonal element to choose (fit to the raw data, the Wiener filtered, etc.).}
\end{enumerate}

\subsection{Assuming no deconvolution of $W({\bf k})$}
\label{subsec:convolved}

The second case is a scenario in which the observation mask was not deconvolved from the underlying field.
This set up introduces many extra challenges, as the mask tends to enhance the non-Gaussian features,
hence, for simplicity, we assume that there is a unique selection function $W({\bf k})$ that covers all the patches in which the power spectra are measured.
Let us first recall that in presence of a selection function, the observed power spectrum of a patch `$i$' is related to the underlying one
via a convolution with the Fourier transform of the mask, namely:
\begin{equation}
P^{i}_{\mbox{obs}}({\bf k}) = \int  P^{i}({\bf k'}) |W({\bf k'} - {\bf k})|^{2}d{\bf k'}
\end{equation}

The first paper of this series describes a general extension to the FKP calculation in which
the underlying covariance matrix $C({\bf k}, {\bf k'})$ is non-Gaussian.
Specifically,  the  `observed' covariance matrix $C_{obs}({\bf k}, {\bf k'})$ is related to the underlying one via a six-dimensional convolution:
\begin{eqnarray}
C_{\mbox{obs}}({\bf k}, {\bf k'})   \propto \sum_{{\bf k''}, {\bf k'''}} C({\bf k''}, {\bf k'''}) \big|W({\bf k} - {\bf k}'')\big|^{2} \big|W({\bf k'} - {\bf k}''')\big|^{2}
\label{eq:obsCov}
\end{eqnarray}

In HDP1, $C({\bf k}, {\bf k'})$ is calculated purely from N-body simulations, therefore it is known {\it a priori};
only $P(k)$ and $W({\bf x})$ are extracted from the survey.
It is in that sense that the technique of HDP1 provides an external estimate of the error.
What needs to be done in internal estimates is to walk these steps backward: given a selection function
and a noisy covariance matrix, how can we extract the non-Gaussian parameters of Table \ref{table:Parameters}?
Ideally, we would like to deconvolve this matrix from the selection function, but the high dimensionality of the integral in [Eq. \ref{eq:obsCov}]
makes the brute force approach  numerically not realistic. 

There is a solution, however, which exploits the fact that many of the terms involved in the forward convolution are linear. 
It is thus possible to perform a least square fit for some of the non-Gaussian parameters, knowing $C_{\mbox{obs}}$ and $W({\bf x})$.
We start by casting the underlying non-Gaussian covariance matrix into its parameterized form,
which expresses each of its Legendre multipole matrices $C_{\ell}^{ij}$ into a diagonal
and a set of Eigenvectors (see  [Eq. 51-54] and Tables 1 and 2 of HDP1\footnote{
In the following discussion, we refer substantially to sections 7.2, 8 and 8.1 of HDP1, and we try to avoid unnecessary repetitions
of lengthy equations here. Also, as discussed in section \protect \ref{subsec:offdiag}, some of the  parameters in Table 2 of HDP1 are degenerate, 
and in fact only 30 are necessary. For each multipole, the main Eigenvector requires only $A$ and $k_{1}$, and we can merge $\lambda$ into $\alpha$ for the others.}). We recall that only the $\ell = 0,2,4$ multipoles contain a significant departure from the Gaussian prescription;
the complete matrix depends on 6 parameters to characterize the diagonal components, 
plus 30 others to characterize the Eigenvectors.
In principle, these 36 parameters could be found all at once by a direct least square fit approach.
However, some of them have more importance than other, and, as seen in this paper, some of them are easier to measure,
therefore we should focus our attention on them first.
In particular, it was shown in figure 22 of HDP1 that most of the non-Gaussian deviations come from $C_{0}$, 
hence we start by solving only for its associated parameters.

To simplify the picture even more, we decompose the problem one step further and focus exclusively on the diagonal component.
In this case, we get, with the notation of the current paper:
\begin{eqnarray}
C(k,k',\theta) \sim \tilde{C}_{0}(k,k', \mu) \equiv C_{G}(k)\delta_{kk'}\bigg( \delta_{\mu 1} + \left(\frac{k}{k_{0}}\right)^{\alpha}\bigg)
\end{eqnarray} 
The `tilde' symbol serves to remind us that this is not the complete $\ell = 0$ multipole but only its diagonal.
The term with the $\delta_{\mu 1}$ is the Gaussian contribution, and yields to ([Eq. 56] of HDP1):
 \begin{eqnarray}
C_{G}^{obs}(k,k')  =  \sum_{k''}C_{G}(k'') \big|W(k'' - k)\big|^{2}\big|W(k''-k')\big|^{2} 
\end{eqnarray}
For the second term, the $\delta_{kk'}$ allows us to get rid of one of the radial integral in [Eq. \ref{eq:obsCov}],
and the remaining part is isotropic, therefore the angular integrals only affect the selection functions.
These integrals can be precomputed as the $X({\bf k}, k'')$ function in [Eq. 57] of HDP1, with $w(\theta'',\phi'') = 1$, and we get
 \begin{eqnarray}
\tilde{C}_{0}^{obs}(k,k')  = C_{G}^{obs}(k,k') +  \sum_{k''}\bigg(\frac{k''}{k_{0}}\bigg)^{\alpha} \bigg\langle X({\bf k}, k'')\bigg\rangle\bigg\langle X({\bf k'}, k'') \bigg\rangle
\label{eq:C_tilde_obs}
\end{eqnarray}
where the angle brackets refer to an average over the angular dependence.
At the end of this calculation, we obtain, for each $(k,k')$ pair, a value for $\alpha$ and $\beta$,
and all that is left is to find the parameter values that minimize the variance. 

%

The next step is to include the off-diagonal terms of the $\ell = 0$ multipole, in which case the full $C_{0}$ is modelled.
The underlying covariance matrix is now parameterized as
 \begin{eqnarray}
C(k,k',\theta) \sim C_{0}(k,k', \mu) \equiv \tilde{C}_{0}(k,k', \mu) + H_{0}(k,k')
\end{eqnarray} 
where 
\begin{eqnarray}
H_{0}(k,k') =   \bigg(F_{0}(k) F_{0}(k') - F_{0}^{2}(k)\delta(k-k')\bigg)
\end{eqnarray}
and 
\begin{eqnarray}
F_{0}(k) = U(k)\sqrt{ \left(1 +   \left( \frac{k}{k_{0}}\right)^{\alpha}\right) C_{G}(k)}
\end{eqnarray}
In this case, [Eq. \ref{eq:C_tilde_obs}] is modified and we get
 \begin{eqnarray}
C_{0}^{obs}(k,k')  = \tilde{C}_{0}^{obs}(k,k') +  \sum_{k'',k''}H_{0}(k'',k''') \bigg\langle X({\bf k}, k'') \bigg\rangle \bigg\langle X({\bf k'}, k''') \bigg\rangle
\end{eqnarray}
There are two new best-fitting parameters that need to be found from $H_{0}$, namely $A$ and $k_{1}$, and we can use our previous results on $\alpha$ and $k_{0}$
as initial values in our parameter search.
Since the $X$ functions only depend on $W({\bf k})$,
we can still solve these $N^2$ equations   with a non-linear least square fit algorithm and extract these four parameters.

Including higher multipoles can be done with the same strategy, i.e. progressively finding new parameters from
least square fits, using the precedent results as priors for the higher dimensional search. 
One should keep in mind that the convolution with $C_{2}$ and $C_{4}$ becomes much more involved, 
since the number of distinct $X$ functions increases rapidly, as seen in Table 4 of HDP1. 
In the end, all of the 36 parameters can be extracted out of $N^2$ matrix elements,
in which case we have fully characterized the non-Gaussian properties of the covariance matrix from the data only.

The matrix obtained this way is expected to have very little noise, as this procedure invokes the fitting functions, which smooth out the fluctuations.
In principle, assuming that the operation was loss-less, the recovered covariance matrix would be completely equivalent to the naive $N=4$ way,
had  the selection function been deconvolved first. 
It could be possible to improve the estimate even further by attempting the T-rotation on the output, as explained is section \ref{subsec:deconvolved},
but since we do not have access to the underlying density fields, the Wiener filter technique is not available in this scenario.





\section{Discussion}
\label{sec:discussion}

With the recent realization that the Gaussian estimator of error bars on the BAO scale is biased  by at least 15 per cent  
\citep{2012MNRAS.419.2949N, 2012MNRAS.423.2288H},
efforts must be placed towards incorporating non-Gaussian features about the matter  power spectrum in data analyses pipeline.
This implies that one needs to estimate accurately the full non-Gaussian covariance matrix $C(k,k')$ and, even more challenging, its inverse.
The strategy of this series of paper is to address this bias issue, via an extension of the FKP formalism that allows for departure from 
Gaussian statistics in the estimate of $C(k,k')$.
The goal of this paper is to develop a method to extract directly from the data the parameters that describe the non-Gaussian features.
This way, the method is free of the biases that affect external non-Gaussian error estimates (wrong cosmology,
incorrect modelling of the non-Gaussian features, etc.).

We emulate a typical observation with a subset of only $N=4$ N-body simulations, and validate our results against a larger $N=200$ sample.
The estimate of $C(k,k')$ obtained with such low statistics is very noisy by nature,
and we develop a series of independent techniques that improve the signal extraction:
\begin{enumerate} 
\item{We break down the full matrix into a diagonal and off-diagonal component, extract the principal component of the latter,
and find best-fitting parameters based on general trends that are known from the $N=200$ sample,}
\item{We write down the full matrix as a sum of a Gaussian component $G(k)\delta_{kk'}$ and 
a non-Gaussian component $NG(k,k')$ and replace the former by the analytical prediction,}
\item{We optimize the uncertainty on the smooth non-Gaussian component from a noise-weighted sum over the different angular contributions,
and apply a Wiener filter on the resulting covariance matrix,}
\item{We rotate the noisy covariance matrix into the Eigenspace of the large sample and weight the elements by the signal-to-noise properties of the Eigenvalues. }
\end{enumerate}

These techniques are exploiting known properties about the noise,  and assume a minimal number of priors about the signal. 
We quantify their performances by comparing their estimate of the  Fisher information content in the matter power spectrum
 to that of  the large  $N=200$ sample. 
 We find that in some cases, we can recover the signal within less then $20$ per cent for $k<1.0h\mbox{Mpc}^{-1}$. 
We also provide error bars about the Fisher information whenever possible.
By comparison, the Gaussian approximation deviates by more than two orders of magnitude at that scale.

We find that the diagonal component of $NG(k,k')$ is well modelled by a simple power law,
and that in the $N=4$ sample, the slope $\alpha$ and the amplitude $k_0$ can be measured with a signal-to-noise ratio of $4.2$ and $2.2$ respectively.
The off-diagonal elements of $NG(k,k')$ are parameterized by fitting the principal Eigenvector of the cross-correlation coefficient matrix 
with two other parameters:  an amplitude $A$, which has a signal-to-noise ratio of $32.1$, and 
a turnaround scale $k_1$, which is measured with a ratio of $0.30$ and is thus the hardest parameter to extract.
Even in the large $N=200$ sample, $k_1$ is measured with a ratio of $3.1$, which is five to ten times smaller than the other parameters.
This help us to better understand the parts of the non-Gaussian signals that are the noisiest, 
such that we can focus our efforts accordingly.

We then propose a strategy to extract these parameters directly from surveys, in the presence of a selection function.
We explore two simple cases, in which a handle of observation patches of equal size, resolution and redshift are combined,
with and without a deconvolution of the survey selection function.
The first case is the easiest to solve, since the deconvolved density fields are in many respect similar 
to the simulated $N=4$ sample that is described in this paper.
We show that it is possible to apply all of the four noise filtering techniques described above
to optimize the estimate of the underlying covariance matrix.

In the second case, we explore what happens when deconvolution is not possible.
We propose a strategy to solve for the non-Gaussian parameters with a least square fit method,
knowing $C_{\mbox{obs}}(k,k')$ and $W({\bf k})$.
The approach is iterative in the sense that it first focuses on the parameters that contribute the most
to the non-Gaussian features, then source the results as priors into more complete searches.

The main missing ingredient from our non-Gaussian parameterization is the inclusion of redshift space distortions, which will be the focus of the next paper of this series.
To give overview of the challenge that faces us, the approach is to expand the redshift space power spectrum
into a Legendre series, and to compute the covariance matrix term by term, again in a low statistics environment.
Namely, we start with $P(k,\mu) = P_{0}(k) + P_{2}(k) \mathcal{P}_{2}(\mu) + P_{4}(k)\mathcal{P}_{4}(\mu) + ...$
where $P_{i}(k)$ are the multipoles and $\mathcal{P}_{\ell}(\mu)$ the Legendre polynomials,
and compute the nine terms in $C(k,\mu,k',\mu') = \langle P_{0}(k) P_{0}(k') \rangle + \langle P_{0}(k) P_{2}(k') \rangle \mathcal{P}_{2}(\mu) + ... $
We see here why a complete analysis needs to include the auto- and cross-correlations between each of these three multipoles,
even without a survey selection function.

As mentioned in the introduction, many other challenges in our quest for optimal and unbiased non-Gaussian error bars
are not resolved yet. 
For instance, many results, including all of the fitting functions from HDP1, were obtained from 
  simulated particles, whereas actual observations are performed from galaxies. It is thus important
  to repeat the analysis with simulated haloes, in order to understand any differences that might exist in the non-Gaussian properties 
  of the two matter tracers. 
In addition, simulations in both HDP1 and in the current paper were performed under a specific cosmology, and only the  $z=0.5$ particle dump
  was analyzed.
 Although we expect higher lower redshift and higher $\Omega_m$ cosmologies to show stronger departures form Gaussianities -- clustering is stronger --
 we do not know how exactly this impact each of the non-Gaussian parameters.



To summarize, we have developed techniques that allow for measurements of non-Gaussian features in the power spectrum uncertainty for galaxy surveys.
We separate the contributions to the total correlation matrix into two kinds: diagonal and off-diagonal.
The off-diagonal is accurately captured in a small number of Eigenmodes, 
and we have used the Eigenframes of the N-body simulations to optimize the measurements
of these off-diagonal non-Gaussian features. 
We show that the rotation into this space allows for an efficient
identification of non-Gaussian features from only four survey fields.    
The method is completely general, and with a large enough sample, we always reproduce the full power spectrum covariance matrix.
We finally describe a strategy to perform such measurements in the presence of survey selection functions.

\section*{Acknowledgments}

The authors would like to thank Chris Blake for reading the manuscript and providing helpful comments concerning
the connections with analyses of galaxy surveys.
UP also acknowledges the financial support provided by the NSERC of Canada. The simulations and computations
were performed on the Sunnyvale cluster at CITA. 

\appendix

\section{Bootstrap vs Noise-weighted integration}
\label{app:noise_weighted}

\begin{figure*}
  \begin{center}
  \epsfig{file=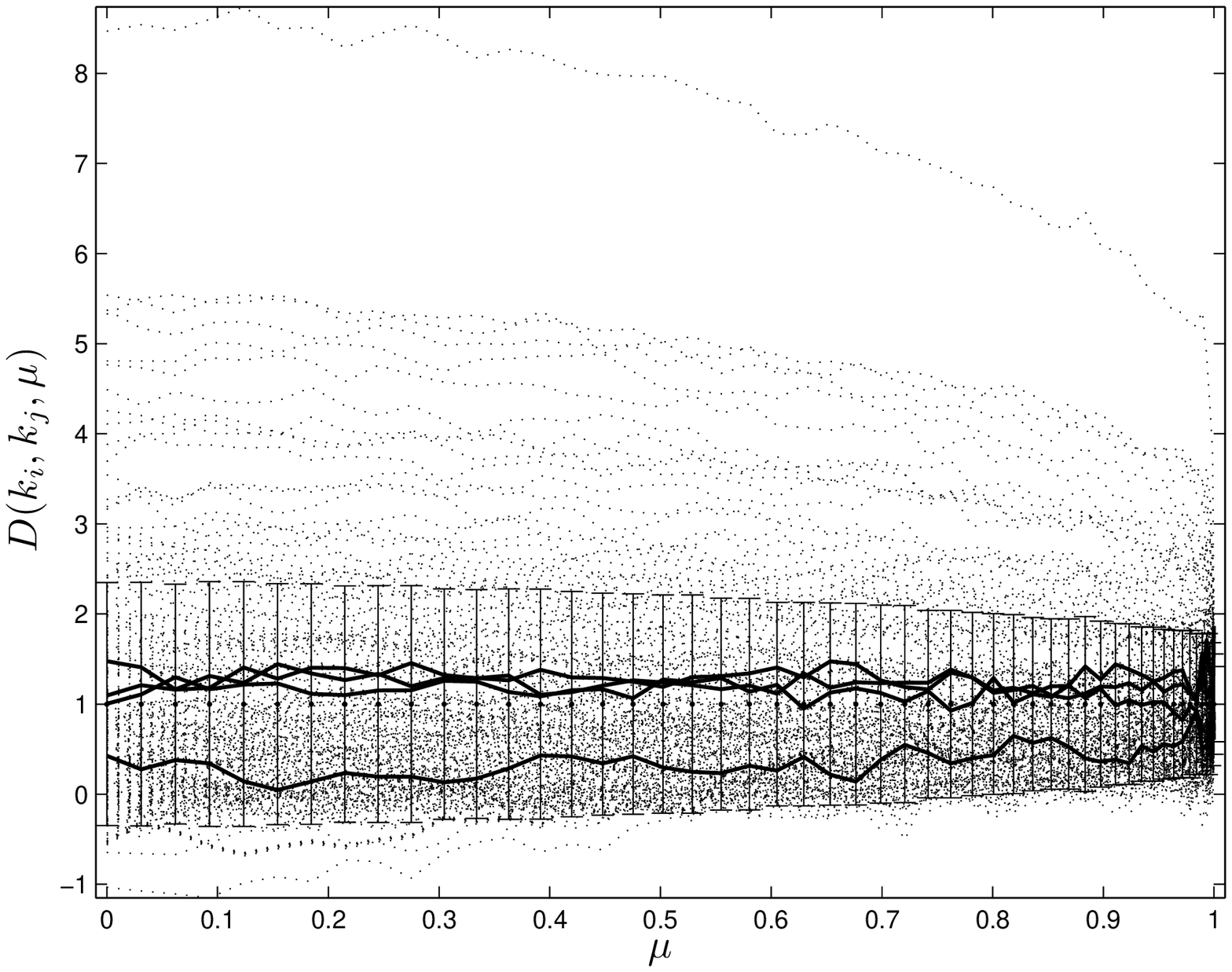,width=0.45\textwidth,height=0.45\textwidth}
  \epsfig{file=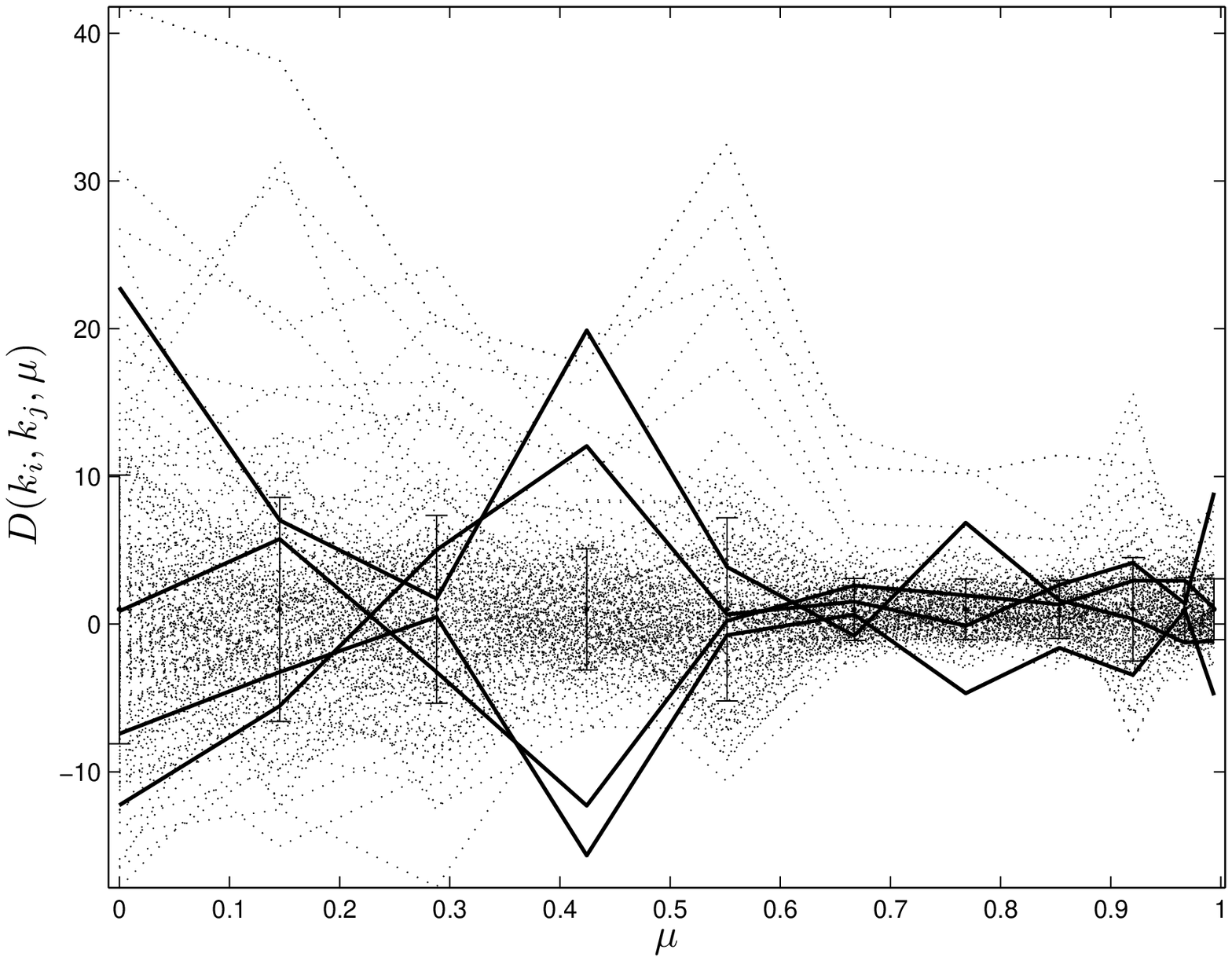,width=0.45\textwidth,height=0.45\textwidth}
  \caption{ ({\it left}:) Fluctuations in the angular dependence of the matter power spectrum covariance, for  $k = k' = 2.36 h \mbox{Mpc}^{-1}$,
  which corresponds to a scale of $2.7 h^{-1}\mbox{Mpc}$. 
  The error bars are the $1\sigma$ standard deviation in the $N=200$ samples. 
  The dotted lines are the individual 200 realizations, while the thin solid lines represent the realizations from the $N=4$ sample. 
  ({\it right}:) Fluctuations for $k = k' = 0.314 h \mbox{Mpc}^{-1}$, 
  which corresponds to a scale of $20.0 h^{-1} \mbox{Mpc}$. 
  The variations in the individual fluctuations are much stronger than in the non-linear regime, 
  a result that approaches the expected behaviour of Gaussian random fields,
  where different measurements are less correlated.}
    \label{fig:D_10_10}
  \end{center}
\end{figure*}

This appendix describes a technique that improves the calculation of the uncertainty on $NG(k,k')$ compared to the bootstrap approach.
Recall that bootstrap combines the error from different angles in quadrature, even though
the signal and noise strength vary for different angles.
To perform the noise-weighted angular integration, we first normalize each realization $C_{i}(k,k',\theta)$
by the mean of the distribution:
\begin{eqnarray}
D_{i}(k,k',\theta) \pm \sigma_{D} \equiv \frac{C_{i}(k,k',\theta)}{C(k,k',\theta)} \pm \frac{\sigma_{C}}{C(k,k',\theta)}
\end{eqnarray}
where $\sigma_{C}$ is the standard deviation in the sampling of $C(k,k',\theta)$. 
As seen in  Fig. \ref{fig:D_10_10}, the individual distributions of $D_{i}(k,k', \theta)$ are relatively flat, with a slight tilt towards smaller angles.
The two panels in this figure correspond to scales $k = k' = 2.1 h \mbox{Mpc}^{-1}$ and 
$k = k' = 0.31 h \mbox{Mpc}^{-1}$ respectively.
In addition, the error bars $\sigma_{D}$ get significantly smaller towards $\theta=0$ (or $\mu = 1$).
It is thus a good approximation to replace each fluctuation by its noise-weighted mean: 
%
\begin{eqnarray}
D_{i}(k,k',\theta) \rightarrow \tilde{D}_{i}(k,k') \equiv \sigma_{T}^{2}\sum_{\theta}\frac{D_{i}(k,k',\theta)}{\sigma_{D}^{2}} \mbox{ with } \sigma_{T}^{-2} = \sum_{\theta} \sigma_{D}^{2}
\label{eq:noiseweightangle}
\end{eqnarray}
The measurement of a matrix element $C_{i}(k,k')$ from a given realization becomes:
\begin{eqnarray}
C_{i}(k,k')&=& G_{i}(k,k') + NG_{i}(k,k') \nonumber \\ 
                 &=& G_{i}(k,k') + \sum_{\theta  \ne 0}C_{i}(k,k', \theta) w(\theta)\nonumber \\
                  &=& G_{i}(k,k') + \sum_{\theta  \ne 0}D_{i}(k,k', \theta) C(k,k',\theta) w(\theta) \nonumber\\
                  &=& G_{i}(k,k') + \tilde{D_{i}}(k,k') \sum_{\theta  \ne 0}C(k,k',\theta) w(\theta)\nonumber\\                  
\end{eqnarray}
where we have used the substitution of [Eq. \ref{eq:noiseweightangle}] in the last step.
Note that in the above expressions, $\sigma_{C,D}$ depend on the variables $(k,k',\theta)$, 
while $\sigma_{T}$ depends on $(k,k')$.
We have chosen not to write these dependencies explicitly in our equations  to alleviate the notation. 
The mean value of $C(k,k')$ computed with this method is identical to the bootstrap approach of [Eq. \ref{eq:integrate}],
since the realization average of $\tilde{D}_{i}(k,k')$ is equal to unity by construction.
However, this method has the direct advantage to reduce significantly the error on $NG$ and, consequently,  on $C$.

\begin{figure*}
  \begin{center}
 \epsfig{file=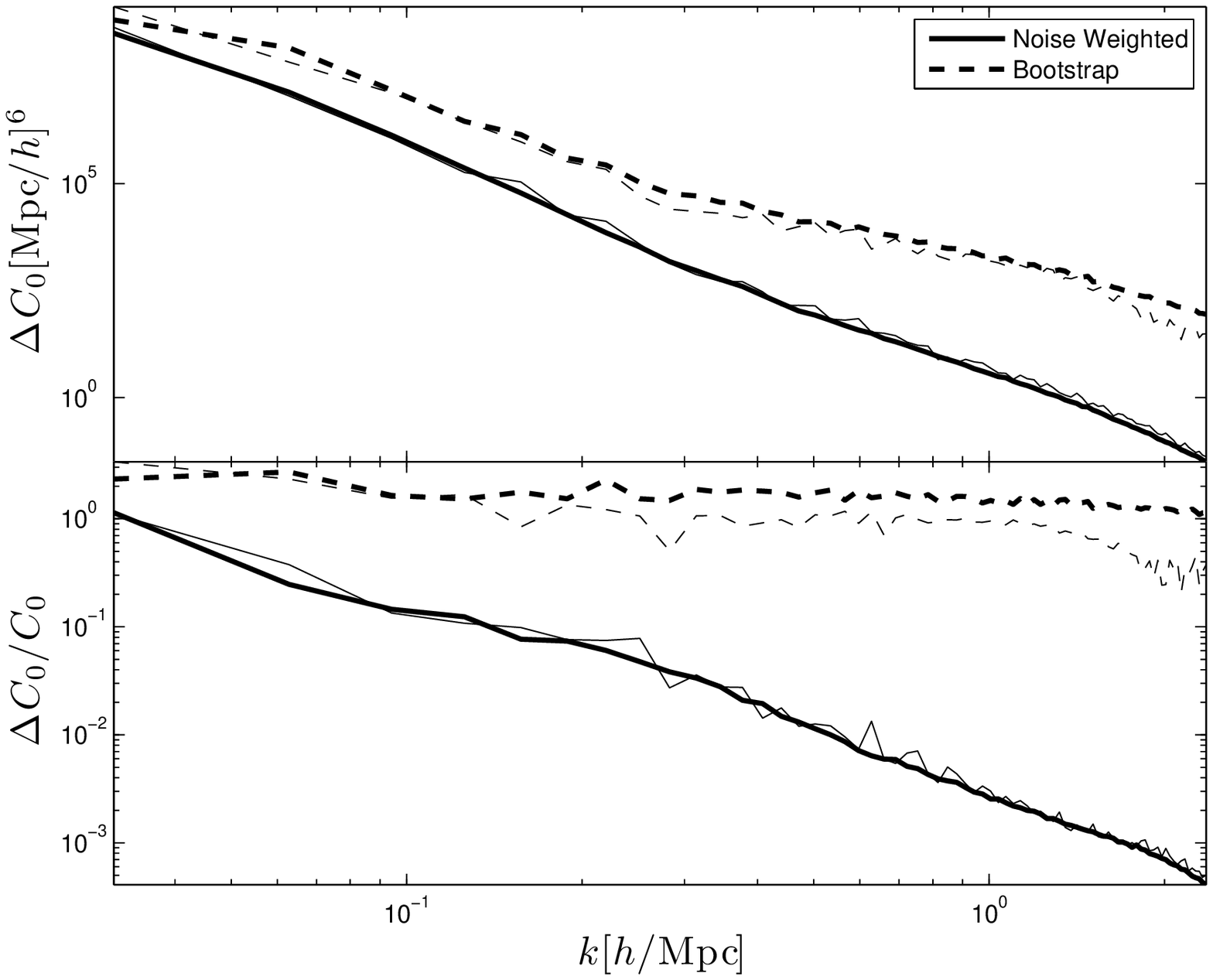,width=0.45\textwidth,height=0.45\textwidth}
    \epsfig{file=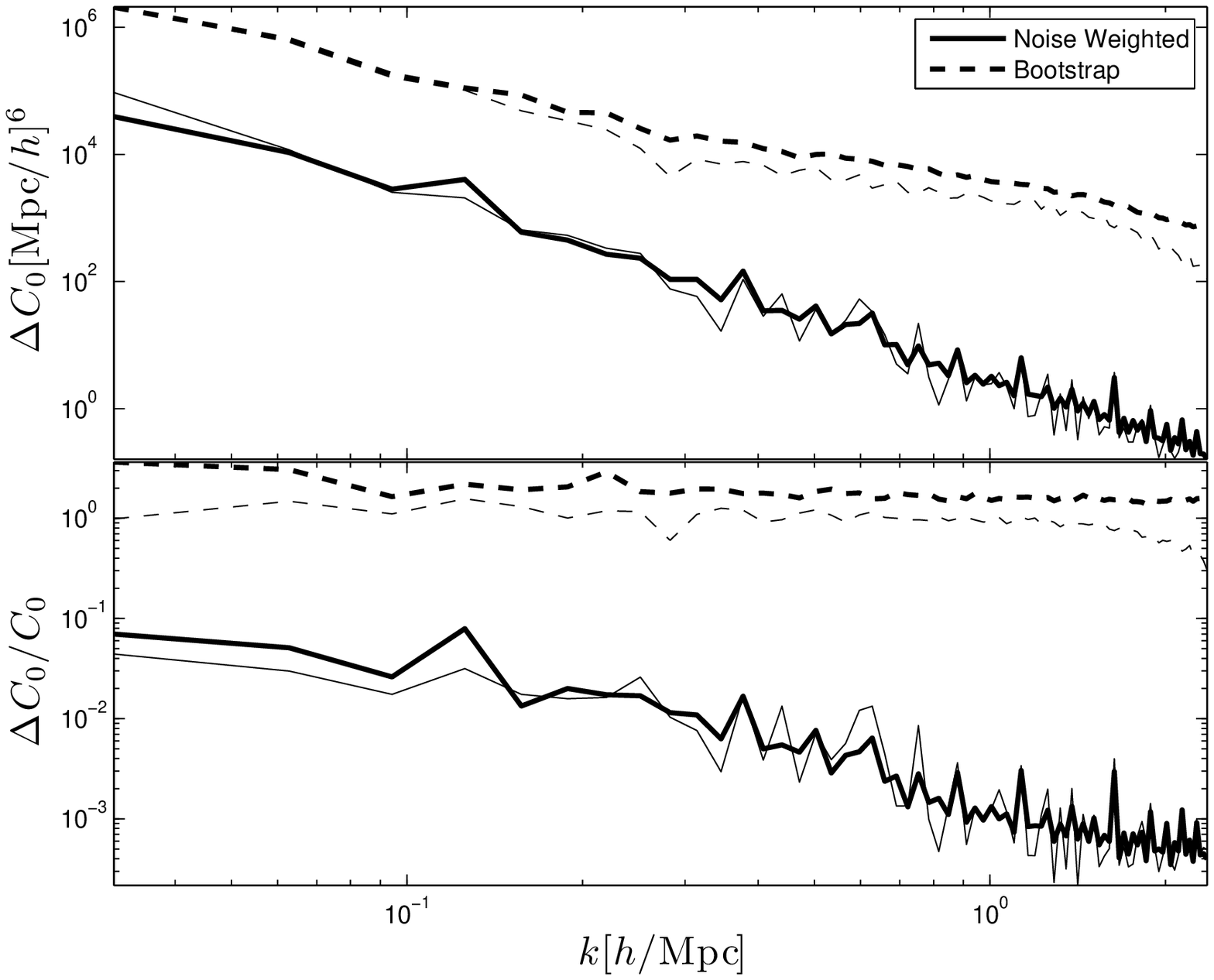,width=0.45\textwidth,height=0.45\textwidth}
  \caption{({\it top left}:) Comparison between bootstrap and noise weighted estimates of the sampling uncertainty on the diagonal of the covariance matrix. 
  We see that the improvement 
  on the precision of this measurement is enhanced significantly in the second scheme, even when 
  very few samples are available. The thick lines are obtained from $200$ realizations, the thin lines from only $4$.
  ({\it bottom left}:) Fractional uncertainty comparison.  
%
  ({\it top and bottom right}:) Same as left column, for the case where one of the scale has been fixed to $ k = 0.628 h \mbox{Mpc}^{-1}$.}
    \label{fig:err_off}
  \end{center}
\end{figure*}

We show in Fig. \ref{fig:err_off} a comparison between the bootstrap sampling error bars on the $C(k,k')$ matrix,
and our proposed noise-weighted scheme. In the left panel, we hold $k' = k$, whereas in the right panel, we keep $k = 0.628 h\mbox{Mpc}$ and vary $k'$.
The error bars achieved  in the noise-weighted scheme are up to two orders of magnitude smaller than the bootstrap errors,
and the estimate of the error from the small sample is already accurate. 
We also observe that the bootstrap fractional error on $C$ is scale invariant, whereas that from the noise-weighted method 
drops roughly as $k^{-2}$ and thereby yields much tighter constraints on the measured matrix elements.
This comes from the fact that as we go to larger $k$-modes, the signal becomes stronger for small angles, which improves the weighting.
In both samples, by the time we have reach  $k=0.1 h \mbox{Mpc}^{-1}$,  the improvement is about 
an order of magnitude, and at $k=1.0 h \mbox{Mpc}^{-1}$, the improvement is almost two orders of magnitude.
With a fractional error that small, the covariance matrix is precisely measured even with a handful of realizations,
a claim that bootstrap approach can not support. 

\bibliographystyle{hapj}
\bibliography{mybib3}{}

\bsp

\label{lastpage}

\end{document}